\RequirePackage{fix-cm}

\documentclass[twocolumn,epjc3]{svjour3}

\RequirePackage[T1]{fontenc}

\RequirePackage{graphicx}
\RequirePackage{subfigure}
\RequirePackage{rotating}
\RequirePackage{pdflscape}
\RequirePackage{color}
\RequirePackage{lineno}
\journalname{Eur. Phys. J. C}

\begin{document}

\title{Development of $^{100}$Mo-containing scintillating bolometers 
for a high-sensitivity neutrinoless double-beta decay search}
\titlerunning{Bolometric technology}

\author{
E.~Armengaud\thanksref{CEA-IRFU}\and 
C.~Augier\thanksref{IPNL}\and 
A.S.~Barabash\thanksref{ITEP}\and 
J.W.~Beeman\thanksref{LBNL}\and 
T.B.~Bekker\thanksref{SIGM}\and 
F.~Bellini\thanksref{Sapienza,INFN-Roma}\and
A.~Beno\^{\i}t\thanksref{Neel}\and 
L.~Berg\'e\thanksref{CSNSM}\and 
T.~Bergmann\thanksref{KIT-IPE}\and 
J.~Billard\thanksref{IPNL}
% ...
}
\thankstext{e1}{e-mail: denys.poda@csnsm.in2p3.fr}
\institute{
IRFU, CEA, Universit\'{e} Paris-Saclay, F-91191 Gif-sur-Yvette, France \label{CEA-IRFU}
\and
Univ Lyon, Universit\'{e} Lyon 1, CNRS/IN2P3, IPN-Lyon, F-69622, Villeurbanne, France \label{IPNL}
\and
National Research Centre Kurchatov Institute, Institute of Theoretical and Experimental Physics, 117218 Moscow, Russia \label{ITEP}
\and
Lawrence Berkeley National Laboratory, Berkeley, California 94720, USA \label{LBNL}
\and
V.S.~Sobolev Institute of Geology and Mineralogy of the Siberian Branch of the RAS, 630090 Novosibirsk, Russia \label{SIGM}
\and
Dipartimento di Fisica, Sapienza Universit\`a di Roma, P.le Aldo Moro 2, I-00185, Rome, Italy \label{Sapienza}
\and
INFN, Sezione di Roma, P.le Aldo Moro 2, I-00185, Rome, Italy \label{INFN-Roma}
\and
CNRS-N\'eel, 38042 Grenoble Cedex 9, France \label{Neel}
\and
CSNSM, Univ. Paris-Sud, CNRS/IN2P3, Universit\'e Paris-Saclay, 91405 Orsay, France \label{CSNSM}
\and
Karlsruhe Institute of Technology, Institut f\"{u}r Prozessdatenverarbeitung und Elektronik, 76021 Karlsruhe, Germany \label{KIT-IPE}
% ...
\and
Karlsruhe Institute of Technology, Institut f\"{u}r Experimentelle Teilchenphysik, 76128 Karlsruhe, Germany \label{KIT-IET}
% ...
\and
INFN, Sezione di Milano Bicocca, I-20126 Milano, Italy \label{INFN-Milano}
\and
IRAMIS, CEA, Universit\'{e} Paris-Saclay, F-91191 Gif-sur-Yvette, France \label{CEA-IRAMIS}
% ...
\and
Orph\'{e}e, CEA, Universit\'{e} Paris-Saclay, F-91191 Gif-sur-Yvette, France \label{CEA-Orphe}
% ...
\and
INFN  Sezione di Genova, I-16146 Genova, Italy \label{INFN-Genova}
\and
Karlsruhe Institute of Technology, Institut f\"{u}r Kernphysik, 76021 Karlsruhe, Germany \label{KIT-IK}
% ...
}
\date{Received: date  / Accepted: date}
\maketitle
\begin{abstract}
This paper reports on the development of a technology involving $^{100}$Mo-enriched scintillating bolometers, compatible with the goals of CUPID, a proposed next-generation bolometric experiment to search for neutrinoless double-beta decay. Large mass ($\sim$1~kg), high optical quality, radiopure $^{100}$Mo-containing zinc and lithium molybdate crystals have been produced and used to develop high performance single detector modules based on 0.2--0.4~kg scintillating bolometers. In particular, the energy resolution of the lithium molybdate detectors near the $Q$-value of the double-beta transition of $^{100}$Mo (3034~keV) is 4--6~keV FWHM. The rejection of the $\alpha$-induced dominant background above 2.6~MeV is better than 8$\sigma$. Less than 10~$\mu$Bq/kg activity of $^{232}$Th ($^{228}$Th) and $^{226}$Ra in the crystals is ensured by boule recrystallization. The potential of $^{100}$Mo-enriched scintillating bolometers to perform high sensitivity double-beta decay searches has been demonstrated with only 10~kg$\times$d exposure: the two neutrino double-beta decay half-life of $^{100}$Mo has been measured with the up-to-date highest accuracy as $T_{1/2}$ = [6.90 $\pm$ 0.15(stat.) $\pm$ 0.37(syst.)] $\times$ 10$^{18}$~yr. Both crystallization and detector technologies favor lithium molybdate, which has been selected for the ongoing construction of the CUPID-0/Mo demonstrator, containing several kg of $^{100}$Mo.
\end{abstract}


%################################################################################################

\section{Introduction}
\label{sec:Intro}

Neutrinoless double-beta ($0\nu 2\beta$) decay, a yet-to-be-ob\-served nuclear transition, consists in the transformation of an even-even nucleus into a lighter isobar containing two more protons with emission of two electrons and no other particles, resulting in a violation of the total lepton number by two units: $(A,Z) \rightarrow (A,Z+2)+2e^-$ (e.g. see Ref. \cite{Vergados:2017}). This hypothetical  transition is energetically allowed for 35 nuclei \cite{Tretyak:2002}. The detection of $0\nu 2\beta$ decay would have profound implications for our understanding of nature, proving that neutrinos are their own antiparticles (Majorana fermions), fixing the absolute neutrino mass scale and offering also a clue for the creation of matter abundance in the primordial universe (see recent reviews \cite{Vergados:2017,DellOro:2016} and references therein). It is to remark that this process is much more than a neutrino-physics experiment, because $0\nu 2\beta$ decay is a powerful, inclusive test of lepton number violation. Non-conservation of the total lepton number is as important as baryon number violation and is naturally incorporated by many theories beyond the Standard Model (SM). The current most stringent lower limits on the $0\nu 2\beta$ decay half-lives are in the range of 10$^{24}$--10$^{26}$~yr \cite{Vergados:2017,Gando:2017}. The SM allowed process two neutrino double-beta ($2\nu 2\beta$) decay is the rarest observed nuclear transition and it has been measured in 11 nuclides with the half-lives in the range of 10$^{18}$--10$^{24}$~yr \cite{Barabash:2015}.

There are a number of proposed next-generation $0\nu 2\beta$ decay experiments, based on upgrades of the most promising current technologies (e.g. see Refs. \cite{Vergados:2017,Cremonesi:2014,Beeman:2012,Beeman:2012a,Artusa:2014a}). The goal of these future searches is to improve by up to two orders of magnitude the present best limits on the half-life with a sensitivity to the effective Majorana neutrino mass (a measure of the absolute neutrino mass scale) at the level of 10--20 meV, covering the so-called inverted hierarchy region of the neutrino mass pattern. The bolometric approach is amongst the most powerful methods to investigate $0\nu 2\beta$ decay. Particularly, one of the most stringent constrains on the effective Majorana neutrino mass \cite{Vergados:2017} have been set by the results of Cuoricino and CUORE-0, precursors of the Cryogenic Underground Observatory for Rare Events (CUORE) \cite{Arnaboldi:2004a}, which studies the candidate isotope $^{130}$Te with the help of TeO$_2$ bolometers. CUORE, a ton-scale $0\nu 2\beta$ decay experiment, is now taking data in the Gran Sasso National Laboratories (Italy) and will be in operation for several years. A large group of interest is proposing a next-generation bolometric experiment, CUORE Upgrade with Particle ID (CUPID) \cite{CUPID,CUPID_RD}, to follow CUORE after the completion of its physics program. The nuclei $^{130}$Te, $^{100}$Mo, $^{82}$Se and $^{116}$Cd are the $0\nu 2\beta$ candidates considered for CUPID. A selection of the CUPID technologies, isotopes and materials is foreseen in 2018/2019. 

Scintillating bolometers, the devices used in the pre\-sent work, are favorable nuclear detectors for the conduction of sensitive $0\nu 2\beta$ decay searches, as they offer high detection efficiency, excellent energy resolution (at the level of $\sim$0.1\%), efficient $\alpha$/$\gamma$($\beta$) particle separation and potentially low intrinsic background \cite{Artusa:2014a,Alessandrello:1992,Bobin:1997,Alessandrello:1998,Meunier:1999,Pirro:2006,Giuliani:2012}. The $^{100}$Mo isotope is one of the most promising $0\nu2\beta$ candidates, since its $0\nu2\beta$ signal is expected at $Q_{\beta\beta}$ = 3034 keV \cite{Rahaman:2008} ($Q$-value of the transition), while the environmental $\gamma$ background mainly ends at 2615 keV. The candidate is embedded in zinc and lithium molybdate crystals (ZnMoO$_4$ and Li$_2$MoO$_4$), working both as low-temperature bolometers and scintillators. An auxiliary bolometer, consisting of a thin Ge wafer, faces each $^{100}$Mo-containing crystal in order to detect the scintillation light. The energy region above $\sim$2.6~MeV is dominated by events produced by radioactive contamination of surfaces, especially $\alpha$ particles (e.g. as shown by the Cuoricino \cite{Andreotti:2011} and CUORE-0 \cite{Alfonso:2015}). Scintillation light yield from alpha interactions is usually quenched when compared to the $\gamma$($\beta$) interactions of the same energy \cite{Tretyak:2010}. Combined with the fact that the thermal response for $\alpha$ and $\gamma$($\beta$) interactions are nearly equivalent, this allows for dual channel scintillating bolometer readouts to perform an effective event-by-event active $\alpha$ background rejection \cite{Beeman:2012,Beeman:2012a,Artusa:2014a,Pirro:2006}. 

The development of a reproducible crystallization and detector technologies is needed for the scintillating bolometer technique to be applicable to a large-scale $0\nu 2\beta$ experiment, like CUPID. The specific requirements to be fulfilled by a crystallization technology of $^{100}$Mo-containing scintillators are \cite{Berge:2014}: large enough crystal boule size; limited losses of the high-cost enriched isotope in the purification-crystallization chain; good optical properties; high scintillation yield; exceptionally low radioactive contamination. The size of a boule should be enough to produce at least one $\sim$70--100~cm$^3$ scintillation element. The volume of the $^{100}$Mo-containing crystal is bounded to the aformentioned value in order to avoid a significant impact on background from random coincidences of the $2\nu 2\beta$ decay events of $^{100}$Mo \cite{Chernyak:2012,Chernyak:2014,Chernyak:2017}. Irrecoverable losses of the enriched material are acceptable at the level of a few \% taking into account that the price of the enriched isotope $^{100}$Mo is $\sim$80~\$/g \cite{Giuliani:2012a}. High transmittance (no less than 30~cm absorption length at the emission maximum) is welcome to reduce the amount of the trapped light and therefore to improve the scintillation light yield \cite{Chernyak:2013}. ZnMoO$_4$ and Li$_2$MoO$_4$ crystals have a reasonable scintillation yield, at the level of 1~keV/MeV, which do not require ultra-low-noise bolometric light detectors. (Baseline noise at the level of a few hundreds of eV are sufficient to provide efficient light-assisted particle identification) According to Monte Carlo simulations of $0\nu 2\beta$ experiments based on $^{100}$Mo-containing scintillating bolometers \cite{Beeman:2012,Beeman:2012a,Artusa:2014a,Danevich:2015,Luqman:2017}, a crystal bulk contamination of the order of 0.01~mBq/kg of $^{228}$Th would result to a minor contribution to the background in the region of interest (ROI; e.g. FWHM wide centered at $Q_{\beta\beta}$), at the level of 10$^{-4}$~counts/yr/kg/keV \cite{Artusa:2014a}. As far as $^{226}$Ra is concerned, a specific activity of even an order of magnitude higher would provide the significantly lower contribution to the background (e.g. see \cite{Danevich:2015,Luqman:2017}). The total activity of other radionuclides from the U/Th chains should not be higher than few mBq/kg to avoid pile-up effects. The main demands concerning the detector performance at the ROI are \cite{Beeman:2012a,Artusa:2014a,Danevich:2015}: better than 10~keV FWHM energy resolution (5~keV FWHM is the CUPID goal \cite{CUPID}); at least 99.9\% rejection of $\alpha$-induced events (with $\gamma$($\beta$)s acceptance larger than 90\%) to suppress this background component to less than 10$^{-4}$~counts/yr/kg/keV. 

Preliminary results have been achieved in the past with bolometers containing molybdenum with natural isotopic composition in the Gran Sasso underground laboratory in Italy \cite{Beeman:2012a,Gironi:2010,Beeman:2012b,Barinova:2010,Cardani:2013}, in the Modane underground laboratory in France \cite{Danevich:2015,Armengaud:2015,Poda:2016,Poda:2015} and in an aboveground cryogenic laboratory located at CSNSM (Orsay, France) \cite{Berge:2014,Chernyak:2013,Beeman:2012c,Mancuso:2014a,Chernyak:2015,Bekker:2016}. In the latter set-up, the first small $^{100}$Mo-enriched ZnMoO$_4$ two-detector array has been tested recently \cite{Barabash:2014}. Most of these R\&D activities were conducted in the framework of the scintillating-bolometer research programs of LUCIFER \cite{lucifer} --- focused on ZnSe for the $0\nu 2\beta$ decay candidate $^{82}$Se but involving also $^{100}$Mo-containing scintillators --- and of LUMINEU \cite{lumineu}, dedicated to the investigation of $^{100}$Mo.
 
The present work represents a crucial step forward in the development of radiopure scintillating bolometers based on ZnMoO$_4$ and Li$_2$MoO$_4$ crystals grown from $^{100}$Mo-enriched molybdenum. A protocol for crystal growth was developed, and several prototypes were tested showing excellent energy resolution, efficient $\alpha$ background rejection power and remarkable radiopurity. The results described here prove in particular that the Li$_2$MoO$_4$ technology is mature enough to carry out a pilot experiment on a several-kilogram scale. This technology demonstrator will provide essential information for the choice of the CUPID technique by clarifying the merits and the drawbacks of the $^{100}$Mo option.  

%################################################################################################

\section{R\&D on natural and $^{100}$Mo-enriched zinc and lithium molybdates}
\label{sec:RnD}

\nopagebreak
\begin{figure*}[htbp]
\centering
  \includegraphics[width=0.4\textwidth]{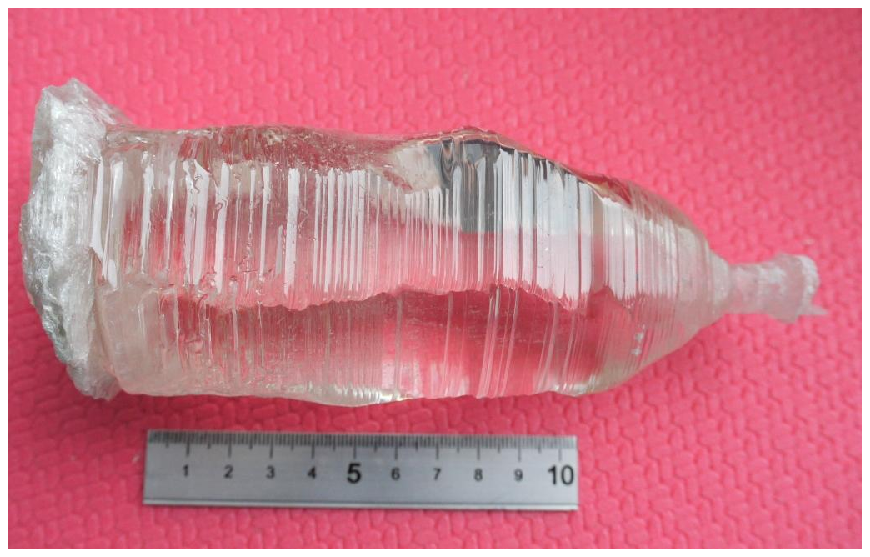}
  \includegraphics[width=0.24\textwidth]{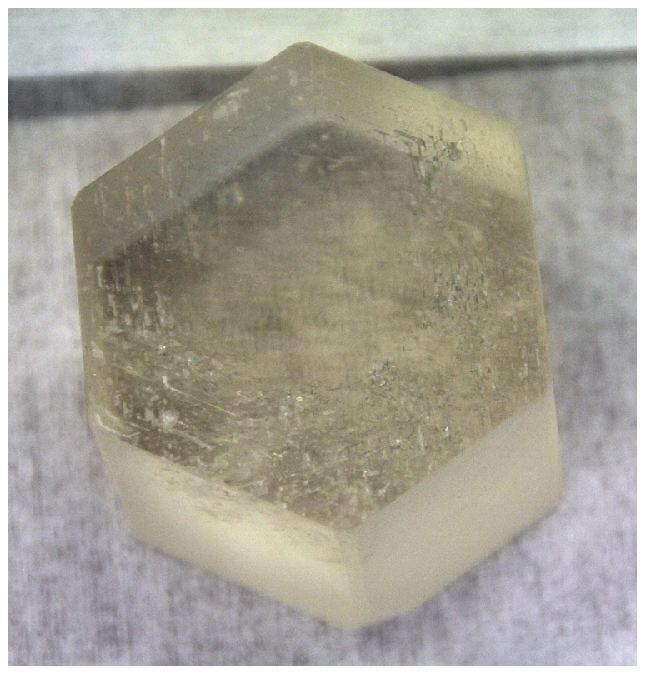}
  \includegraphics[width=0.4\textwidth]{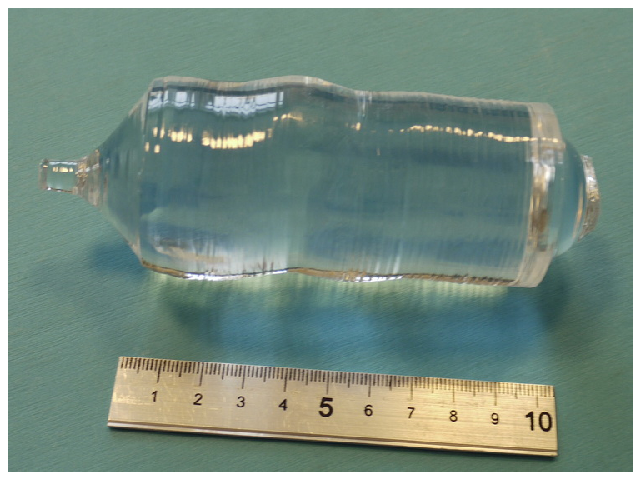}
  \includegraphics[width=0.24\textwidth]{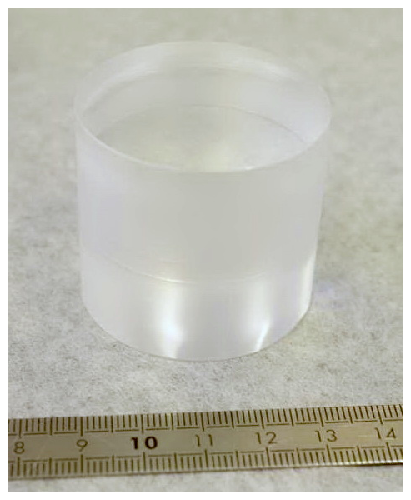}
	\caption{Photographs of the first large-mass $^{100}$Mo-enriched scintillators: the $\sim$1.4~kg  Zn$^{100}$MoO$_4$ crystal boule with the cut $\sim$0.38~kg scintillation element enrZMO-t (top panels), and the $\sim$0.5~kg boule of Li$_2$$^{100}$MoO$_4$ crystal with the produced $\sim$0.2~kg sample enrLMO-t (bottom panels). Both scintillation elements were cut from the top part of the boules. Color and transparency of the enrZMO-t crystal are different from the ones of the boule due to artificial light source and grinded side surface. The photo on the top left panel is reprinted from \cite{Poda:2015}}
\label{fig:Crystals}
\end{figure*} 

Important milestones were achieved by LUMINEU in the R\&D on zinc molybdate scintillators: the development of a molybdenum purification procedure \cite{Berge:2014}; the growth of large ($\sim$1~kg) ZnMoO$_4$ \cite{Armengaud:2015} and small (0.17~kg) Zn$^{100}$MoO$_4$ \cite{Barabash:2014} crystals with the help of the low-tempe\-rature-gradient Czochralski (LTG Cz) method \cite{Pavlyuk:1992,Borovlev:2001}; further optimization of the ZnMoO$_4$ growth process \cite{Chernyak:2015}. The R\&D goal has been accomplished by the successful development of a large-mass Zn$^{100}$MoO$_4$ crystal boule ($\sim$1.4~kg in weight, $^{100}$Mo enrichment is $\sim$~99\%) shown in Fig.~\ref{fig:Crystals} (top left). Even though there is still room to improve the Zn$^{100}$MoO$_4$ crystal quality --- the boule exhibits a faceted structure and contains inclusions, mainly in the bottom part --- the developed Zn$^{100}$MoO$_4$ crystallization technology ensures the growth of reasonably good quality scintillators with a mass of about 1~kg --- which represents more than 80\% yield from the initial charge of the powder in the crucible --- and below 4\% irrecoverable losses of the enriched material. 

In the present work, we report about the study of four massive (0.3--0.4~kg) ZnMoO$_4$ crystals operated as scintillating bolometers at $\sim$(10--20)~mK. Two scintillation elements have been cut from a boule containing molybdenum of natural isotopic composition \cite{Armengaud:2015}, while the other two are obtained from the Zn$^{100}$MoO$_4$ boule (Fig.~\ref{fig:Crystals}, top left). The information about the applied molybdenum purification, the size and the mass of the produced samples are listed in Table \ref{tab:crystals}. The size of crystals is chosen to minimize the material losses and to produce similar-size samples from each boule. According to \cite{Danevich:2014}, a hexagonal shape of Zn$^{100}$MoO$_4$ elements (e.g. see Fig.~\ref{fig:Crystals}, top right) should provide a higher light output than the cylindrical one.  

\begin{table*}[htb]
\caption{Zinc and lithium molybdate crystal scintillators grown by the LTG Cz method from molybdenum with natural isotopic composition and enriched in $^{100}$Mo. The molybdenum compound has been purified by single or double sublimation with subsequent double recrystallization in aqueous solutions. A Li$_2$CO$_3$ compound supplied by NRMP (see text) was used to produce all Li-containing scintillators, except LMO-3 (produced from Alfa Aesar Li$_2$CO$_3$). The position in the crystal boule is given for those samples cut from the same boule. The crystal ID is represented by the abbreviation of the chemical compound with an extra ``enr'' to mark enriched samples and/or ``t'' or ``b'' to indicate the position in the boule and a number to distinguish boules of the same material}
\scriptsize
\begin{center}
\begin{tabular}{|c|c|c|c|c|c|c|}
 \hline
\bf{Scintillator}			& \bf{Molybdenum}		& \bf{Boule}						& \bf{Crystal}	& \bf{Position}	& \bf{Size}											& \bf{Mass}	 \\
~             				& \bf{sublimation} 	& \bf{crystallization}	& \bf{ID}				& \bf{in boule}	& \bf{($\oslash$$\times$h mm)} 	& \bf{(g)}	\\
\hline
ZnMoO$_4$							& Single	& Double	& ZMO-t			& top				& 50$\times$40	& 336	 \\
~             				& ~				& ~ 			& ZMO-b			& bottom		& 50$\times$40	& 334	\\
\hline
Zn$^{100}$MoO$_4$			& Double	& Single	& enrZMO-t	& top				& 60$\times$40	& 379	 \\
~											& ~				& ~				& enrZMO-b	& bottom		& 60$\times$40	& 382	 \\
\hline
Li$_2$MoO$_4$					& Single	& Single	& LMO-1			& --				& 40$\times$40	& 151	 \\
\cline{2-7}
~											& Single	& Double	& LMO-2			& --				& 50$\times$40	& 241	 \\
\cline{2-7}
~											& Single	& Single	& LMO-3			& --				& 50$\times$40	& 242	 \\
\hline
Li$_2$$^{100}$MoO$_4$	& Double	& Triple	& enrLMO-t	& top				& 44$\times$40	& 186	 \\
~											& ~				& ~				& enrLMO-b	& bottom		& 44$\times$44	& 204	 \\
\cline{2-7}
~											& Double	& Double	& enrLMO-2t	& top				& 44$\times$46	& 213	 \\
~											& ~				& ~				& enrLMO-2b	& bottom		& 44$\times$44	& 207	 \\

\hline
\end{tabular}
\end{center}
\label{tab:crystals}
\end{table*}
\normalsize

Because of some experienced difficulties with the ZnMoO$_4$ crystallization process\footnote{We observed the formation of a second phase due to an unstable melt in the ZnO-MoO$_3$ system \cite{Chernyak:2015}.}, which prevented us from obtaining top quality large-mass crystals, the LUMINEU collaboration initiated an R\&D on the production of large-mass radiopure lithium molybdate scintillators \cite{Bekker:2016,Velazquez:2017}. Thanks to the low and congruent melting point of Li$_2$MoO$_4$, the growth process is expected to be comparatively easier than that of ZnMoO$_4$. However, the chemical affinity of lithium and potassium results in a considerably high contamination of $^{40}$K ($\sim$0.1~Bq/kg) in Li$_2$MoO$_4$ crystal scintillators, as it was observed in early studies of this material \cite{Barinova:2009}. Despite the low $Q_{\beta}$ of $^{40}$K, random coincidences of $^{40}$K and $2\nu2\beta$ events of $^{100}$Mo can produce background in the ROI \cite{Chernyak:2014}. In particular, a contamination level around $\sim$0.06~Bq/kg of $^{40}$K in a Li$_2$$^{100}$MoO$_4$ detector with dimension $\oslash50\times40$~mm provides the same background counting rate in the ROI as the random coincidences of the $2\nu2\beta$ events. So, in addition to the LUMINEU specifications on U/Th contamination, the acceptable $^{40}$K activity in Li$_2$MoO$_4$ crystals is of the order of a few mBq/kg. Therefore, the R\&D on Li$_2$MoO$_4$ scintillators included the radioactive screening and selection of commercial lithium carbonate samples, the optimization of the LTG Cz crystal growth and the investigation of the segregation of radioactive elements in the crystallization process. 

Three samples of high purity lithium carbonate were measured by HPGe spectrometry at the STELLA facility of the Gran Sasso underground laboratory (Italy): 
1) 99.99\% purity grade powder produced by Novosibirsk Rare Metal Plant (NRMP, Novosibirsk, Russia) \cite{NRMP}; 
2) 99.998\% lithium carbonate by Puratronic (Alfa Aesar GmbH \& Co KG, France) \cite{Alfa-Aesar}; 
3) 99.99\% raw material by Sigma-Aldrich (USA) \cite{Sigma-Aldrich}. 
The results are given in Table \ref{tab:Li_purity}. The lithium carbonate produced by NRMP, the material of highest radiopurity, was selected for Li$_2$MoO$_4$ crystals production. Due to the high $^{40}$K contamination, the Sigma-Aldrich material was rejected for further investigation. 

\begin{table}[htb]
\caption{Radioactive contamination of commercial lithium carbonate compounds measured by low background HPGe $\gamma$ spectrometry. Errors are given in parenthesis at 68\% C.L., upper limits --- at 95\% C.L.}
\scriptsize
\begin{center}
\begin{tabular}{|c|c|ccc|}
 \hline
\bf{Chain} 	& \bf{Nuclide} 	& \multicolumn{3}{c|}{\bf{Activity in Li$_2$CO$_3$ powder (mBq/kg)}} \\
\cline{3-5}
~          	&  ~           	& NRMP 			& Alfa Aesar	& Sigma-Aldrich \\
\hline
$^{232}$Th 	& $^{228}$Ra 		& $\leq2.9$ & $\leq14$ 		& 16(8)  				\\
~          	& $^{228}$Th 		& $\leq3.7$ & 12(4)				& 13(4)				  \\
\hline
$^{238}$U 	& $^{226}$Ra 		& $\leq3.3$ & 705(30) 		& 53(6)  			  \\
\hline
~          	& $^{40}$K   		& $\leq42$	& $\leq42$ 		& 210(70) 		 \\
 \hline
\end{tabular}
\end{center}
\label{tab:Li_purity}
\end{table}
\normalsize

Even first attempts of the Li$_2$MoO$_4$ growth by the LTG Cz technique were successful providing high quality crystal boules with masses of 0.1--0.4~kg  \cite{Bekker:2016}. The growing conditions 
have been optimized to extend the Li$_2$MoO$_4$ crystal size up to 100~mm in length and 55~mm in diameter (0.5--0.6~kg mass) \cite{Grigorieva:2017} allowing us to produce two large scintillating elements of about 0.2~kg each from one boule. For the present study, we developed three large Li$_2$MoO$_4$ scintillators by using highly purified molybdenum oxide and high purity grade lithium carbonate. Two of them have been grown from the NRMP Li$_2$CO$_3$ compound by applying a single (LMO-1 sample in Table \ref{tab:crystals}) and a double (LMO-2) crystallization, while the last one (LMO-3) was grown by the single crystallization from the Alfa Aesar Li-containing powder.

\begin{table*}[htb]
\caption{The main construction elements of $^{100}$Mo-containing heat detectors studied in the present work. Their IDs coincide with the scintillation crystal IDs defined above. 
Three types of reflectors were used: Radiant Mirror Film (RMF) VM2000/VM2002 and Enhanced Specular Reflector (ESR) film by 3M\texttrademark and a thin silver layer (Ag) deposited 
on the holder. The masses of all used NTD sensors are $\approx$~50~mg. The enrZMO-t, enrZMO-b, and LMO-2 detectors were also equipped with a smeared $^{238}$U $\alpha$ source}
\scriptsize
\begin{center}
\begin{tabular}{|c|c|c|c|c|c|c|}
 \hline
\bf{Standard}	& \bf{Heat}		& \multicolumn{2}{c|}{\bf{Support}}  & \bf{Reflector}	& \multicolumn{2}{c|}{\bf{NTD sensor type}}  \\
\cline{3-4}
\cline{6-7}
~					& \bf{detector ID}	& Copper & 	PTFE 			& ~ 	& No.1	& No.2		\\
\hline
LUMINEU		& ZMO-t			& Holder	& L- and S-	& RMF	& HR	& HR	 	\\
~         & ZMO-b			& ~				& shaped		& ~		& HR	& LR		\\
~					& LMO-1			& ~				& ~					& ~		& LR	& --		\\
~					& LMO-3			& ~				& ~					& Ag	& LR	& --	 	\\
\hline
LUMINEU   & ZMO-b			& Holder	& L- and S-	& ESR	& LR	& --	 	\\
(tower)   & enrLMO-t	& ~				& shaped		& ~		& LR	& --	  \\
\hline
LUCIFER		& enrZMO-t	& Plate,	& S-shaped	& ESR	& LR	& LR	 	\\
~					& enrZMO-b	& columns	& ~					& ~		& LR	& LR	 	\\
~					& LMO-2			& ~				& ~					& ~		& LR	& LR		\\
~					& enrLMO-b	& ~				& ~					& ~		& HR	& HR	 	\\
\hline
\end{tabular}
\end{center}
\label{tab:Bolo}
\end{table*}
\normalsize

Once the LTG Cz growth of Li$_2$MoO$_4$ crystals containing molybdenum of natural isotopic composition was established, we started to process molybdenum enriched in $^{100}$Mo. Fig. \ref{fig:Crystals} (bottom left) shows a first large-mass ($\sim$0.5~kg) $^{100}$Mo-enriched crystal boule grown at the beginning of 2016. The crystal was produced by a triple crystallization due to an accident that happened during the second crystal growth process. The second massive Li$_2$$^{100}$MoO$_4$ crystal boule ($\sim$0.6~kg; see Fig.~4 in \cite{Grigorieva:2017}) was grown by double crystallization at the end of May 2016. Both enriched crystals demonstrated high optical quality and have the size required for the production of two similar transparent Li$_2$$^{100}$MoO$_4$ scintillation elements with masses of $\sim$0.2~kg (see Table \ref{tab:crystals} and Fig. \ref{fig:Crystals}, bottom right). Two cylindrical samples produced from the first Li$_2$$^{100}$MoO$_4$ crystal boule were used for the bolometric tests described in the present work\footnote{All these Li$_2$$^{100}$MoO$_4$ elements have been recently operated as a four-bolometer array in the EDELWEISS set-up at Modane Underground Laboratory (France) \cite{Poda:2017}.}.

%################################################################################################
\section{Underground tests of $^{100}$Mo-containing scintillating bolometers}

%-------------------------------------------------------------------------------------------------
\subsection{$^{100}$Mo-containing scintillating bolometers}
\label{sec:Bolometers}

The bolometers were fabricated from the crystal scintillators listed in Table \ref{tab:crystals}. Each scintillating crystal was equipped with one or two epoxy-glued Neutron Transmutation Doped (NTD) 
Ge temperature sensors \cite{Haller:1994}, whose resistance exponentially depends on temperature as $R(T) = R_{0} \cdot \mathrm{exp}(T_{0}/T)^{\gamma}$. $R_0$ and $T_0$ are two parameters depending on the doping, the compensation level and on the geometry in the case of $R_0$. In our samples, ${\gamma}$ is derived to be 0.5. In the present work we used high resistance (HR) and low resistance (LR) sensors with typical parameter values $T_0$ = 4.8~K, $R_0$ = 2.2~$\mathrm{\Omega}$ and $T_0$ = 3.9~K, $R_0$ = 1.0~$\mathrm{\Omega}$, respectively. Therefore, HR NTDs have a resistance of $\sim$10~M$\mathrm{\Omega}$ at $\sim$20~mK working temperature, while an order of magnitude lower resistance is typical for LR NTDs. The NTD Ge thermistors, biased with a constant current, act as temperature-voltage transducers. The thermal link to the bath was provided by Au bonding wires which give also the electrical connection with the NTD Ge sensors. In addition, each crystal was supplied with a small heater made of a heavily-doped Si \cite{Andreotti:2012}, through which a constant Joule power can be periodically injected by a pulser system to stabilize the bolometer response over temperature fluctuations \cite{Andreotti:2012,Alessandrello:1998b}. 

\begin{figure*}[htbp]
\centering
  \includegraphics[width=0.75\textwidth]{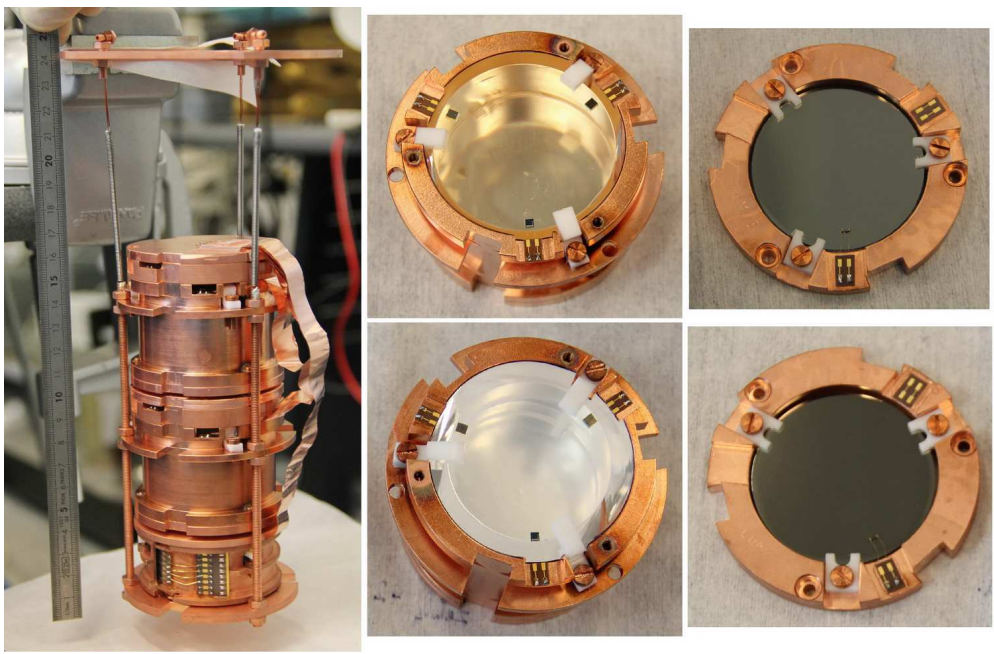}
\caption{Photographs of a three-spring suspended tower (first column) and two LUMINEU scintillating bolometers (second column): the 334~g ZnMoO$_4$ (top photo; ZMO-b, $\oslash 50 \times 40$~mm) and the 186~g Li$_2$$^{100}$MoO$_4$ (bottom photo; enrLMO-t, $\oslash 44 \times 40$~mm) bolometers together with two identical $\oslash$44-mm Ge light detectors (third column). A third detector of the tower (bottom in the left photo, not shown in details) is a 0.2~kg Ge bolometer}
\label{fig:tower}
\end{figure*}

The detectors were assembled according to either LUMINEU or LUCIFER standard schemes (see Table~\ref{tab:Bolo}). The mechanical structure and the optical coupling to the crystal scintillators are designed to optimize the heat flow through the sensors and to maximize the light collection. The standard adopted by LUMINEU for the EDELWEISS-III set-up implies the use of a dedicated copper holder where the crystal scintillator is fixed by means of L- and S-shaped PTFE clamps \cite{Armengaud:2015,Poda:2016,Poda:2015}. The holder is completely covered internally by a reflector to improve the scintillation-light collection. For the prototype of the LUMINEU suspended tower, shown in Fig. \ref{fig:tower}, the holders were slightly modified to make the array structure able to pass through the holes in the copper plates of the EDELWEISS set-up. In case of the LUCIFER R\&D standard, the crystal is fixed to a copper frame by S-shaped PTFE pieces and copper columns, as well as side-surrounded by a plastic reflective film (e.g. see in \cite{Beeman:2012a}). This frame is thermally anchored to the mixing chamber of the dilution refrigerator.

Thin bolometric light detectors (see Table \ref{tab:LD}) were coupled to the scintillating crystals to register the scintillation light. All of them are based on high purity Ge wafers and their typical size is 44--45~mm in diameter and 0.17--0.3~mm thickness, but two detectors have slightly lower area and tens $\mu$m thickness. Some light detectors were constructed according to the LUMINEU standard described in \cite{Tenconi:2015a}, with the additional deposition of a 70~nm SiO antireflecting coating on one surface of the Ge wafer to increase the light absorption \cite{Mancuso:2014}. Another type of light detectors used in the present study was developed by the LUCIFER group \cite{Beeman:2013b}. One bolometer was assembled according to CUPID-0 mounting standard \cite{Artusa:2016}. In all these cases, the Ge wafer is held by PTFE clamps. The last type of used light detectors is the state-of-the-art optical bolometer developed at IAS (Orsay, France) \cite{Coron:2004}. The suspension of the Ge wafer is carried out by Nb-Ti wires in this case. All the light detectors were equipped with one NTD Ge thermistor.

\nopagebreak

\begin{table*}[htb]
\caption{Information about Ge light detectors used in the present work. The detectors are grouped according to the mounting standard given in the first column}
\scriptsize
\begin{center}
\begin{tabular}{|c|c|c|c|c|c|c|}
 \hline
\bf{Standard}	& \bf{Light}				& \bf{Ge size (mm)} 					& \bf{Coating}	& \multicolumn{2}{c|}{\bf{NTD sensor}}  \\
\cline{5-6}
~					& \bf{detector ID}	& ~											& ~ 			& Type	& Mass (mg)	\\
\hline
LUMINEU		& M1			& $\oslash$44$\times$0.17		& yes			& LR	& 20 	\\
~         & M3			& $\oslash$44$\times$0.17		& yes			& LR	& 9 	\\
~         & Lum11		& $\oslash$44$\times$0.17		& yes			& HR	& 5 	\\
~         & Lum12		& $\oslash$44$\times$0.17		& yes			& HR	& 5  \\
\hline
LUCIFER		& GeB			& $\oslash$45$\times$0.30		& no			& LR	& 9 	\\
~					& GeT			& $\oslash$45$\times$0.30		& no			& LR	& 9 	\\
\hline
CUPID-0		& GeOld		& $\oslash$45$\times$0.30		& no			& LR	& 9 	\\
\hline
IAS				& B297		& $\oslash$40$\times$0.045	& no			& LR	& 1 	\\
~					& B304		& $\oslash$25$\times$0.030	& no			& LR	& 1 	\\
\hline
\end{tabular}
\end{center}
\label{tab:LD}
\end{table*}
\normalsize

%-------------------------------------------------------------------------------------------------
\subsection{Underground cryogenic facilities}
\label{sec:Facilities}

In the present investigations, we used two cryogenic set-ups: CUPID R\&D and EDELWEISS-III located at Gran Sasso National Laboratories (LNGS, Italy) and Modane underground laboratory (LSM, France), respectively. The general description of these facilities is given in Table \ref{tab:setup}. Some features are related to the specific applications: the CUPID R\&D is mainly oriented on the R\&D of bolometers (including scintillating bolometers) for $0\nu2\beta$ searches, with ROI at a few MeV, while the EDELWEISS-III set-up was conceived to perform direct dark-matter searches with the help of massive heat-ionization bolometers, with a ROI in the tens-of-keV range.

\begin{table*}[htb]
\caption{A short description of the used underground cryogenic set-ups. The rock overburden is expressed in km of water equivalent (km w.e.). The base temperature indicates the minimal temperature of the cryostat. The sampling rate is given in kilo-samples per sec (kSPS )}
\scriptsize
\begin{center}
\begin{tabular}{|c|c|c|c|}
 \hline
\multicolumn{2}{|c|}{\bf{~}} 	& \bf{CUPID R\&D} 	& \bf{EDELWEISS-III} \\
\multicolumn{2}{|c|}{\bf{~}} 	& \cite{Pirro:2006b,Arnaboldi:2006,Arnaboldi:2004} 	& \cite{Armengaud:2015,Hehn:2016,EDW-performance} \\
\hline
\bf{Location} & Underground lab      & LNGS (Italy)		& LSM (France) 		 \\
~ & Rock overburden (km w.e.)	& 3.6							& 4.8 \\
\hline
\bf{Cryostat} & Dilution refrigerator & $^3$He/$^4$He & $^3$He/$^4$He \\
~ & Type & wet & wet and dry \\
~ & Geometry & standard & reversed \\	
~ & Experimental volume (L) & $\sim$8	& $\sim$50 \\
~ & Outside mechanical decoupling 	& no & yes  \\		
~ & Inside mechanical decoupling 	& yes & yes (since 2016) \\		
~ & Base temperature (mK) & 7	& 10 \\
\hline
\bf{Shield} & Low activity lead (cm) & 20 & 18 \\
\bf{(external)} & Roman lead (cm) & no & 2 \\
~ & Polyethylene (cm) & 8 & 55 \\
~ & Boron carbide (cm) & 1 & no \\
~ & Anti-radon box & yes & no \\
~ & Muon veto & no & yes \\
\hline
\bf{Shield} & Roman lead (cm) & 5.5 & 14 \\
\bf{(internal)} & Polyethylene (cm) & no & 10 \\
\hline
\bf{Readout} & Electronics & Cold + Room-Temp. & Cold \\
\bf{and DAQ} & Dual readout channels & 10 + 8 & 48 \\
~ & Bias & DC & AC \\
~ & ADC digitization (bit) & 18 & 16 or 14 \\
~ & Sampling rate (kSPS) & up to 250/N$_{channels}$ & up to 1 \\
~ & Data taking mode & trigger and/or stream & trigger or stream \\
\hline
\bf{Calibration} & Regular & $^{232}$Th & $^{133}$Ba \\
~ & Exceptional & $^{40}$K, $^{137}$Cs, AmBe & $^{232}$Th, $^{40}$K, AmBe \\
~ & In-situ source  & allowed & prohibited \\
~ & Pulser system & yes & yes (since 2015) \\
 \hline
\end{tabular}
\end{center}
\label{tab:setup}
\end{table*}
\normalsize

As one can see from Table \ref{tab:setup}, an efficient suppression of the cosmic-ray flux is provided by a deep underground location of both set-ups. The EDELWEISS-III is larger and can host up to 48 scintillating bolometers with a copper holder size of $\approx \oslash$80$\times$60~mm each. The reversed geometry of the EDELWEISS-III cryostat does not allow to decouple mechanically the detectors plate from the mixing chamber, as it was done by two-stage damping system inside the CUPID R\&D set-up \cite{Pirro:2006b}. The external damping system (pneumatic dampers) of the EDELWEISS-III is adapted to the operation of tightly held massive EDELWEISS detectors, and not to scintillating bolometers. In particular, thin light detectors are very sensitive to the vibrations induced by the three thermal machines of the set-up. Therefore, an internal damping inside the EDELWEISS-III has been implemented through a mechanically-isolated suspended tower (see Fig. \ref{fig:tower}). Dilution refrigerators of both set-ups are able to reach a base temperature around 10 mK. 

The EDELWEISS-III set-up is surrounded by a significantly massive passive shield against gamma and neutron background. The absence of an anti-radon system as that used in the CUPID R\&D is somehow compensated by a deradonized (below 20~mBq/m$^3$) air flow. The radon level is monitored continuously. An important advantage of the EDELWEISS-III set-up is a muon veto system with about 98\% coverage (however, no clock synchronization with scintillating bolometers has been implemented yet).
 
All the EDELWEISS-III readout channels utilize a cold electronics stage, while only about half of those of CUPID R\&D have this feature. The EDELWEISS-III readout system uses AC bolometer bias modulated at a frequency of up to 1~kHz, which is also kept for the demodulation procedure applied to the data sampled with a 100~kSPS rate (the modulated data can be also saved). Higher resolution without a significant enlargement of the data size is available for the CUPID R\&D case, which envisages DC bolometer bias. In contrast to DC current, there are difficulties in the operation of high resistance NTDs with AC bias (e.g. unbalanced compensation of nonlinearities related to the differentiated triangular wave applied for NTD excitation --- see details in \cite{EDW-performance,Gaertner:1997}).

\begin{table*}[htb]
\caption{General information about measurements with $^{100}$Mo-containing bolometers operated at Modane and Gran Sasso underground laboratories. IDs of detectors used for the construction of double read-out hybrid bolometers correspond to the heat and light detectors ID defined above. $T_{base}$ denotes the base temperature of the cryostat}
\scriptsize
\begin{center}
\begin{tabular}{|c|c|c|c|c|c|c|c|}
 \hline
\bf{Set-up}					& \bf{Run ID}	& \multicolumn{2}{c|}{\bf{Detectors}}	& \multicolumn{2}{c|}{\bf{Sampling (kSPS)}}	& \bf{$T_{base}$} 	& \bf{Data} \\
\cline{3-6}
~								& ~	& Heat	& Light	& Heat & Light	& \bf{(mK)}	& \bf{taking (h)}  \\
\hline
EDELWEISS-III		& Run308	& ZMO-t				& M1		& 1 	& 1 	& 18 			& 5000 	 	\\
~             	& ~				& ZMO-b				& M3		& ~		& ~		& ~				& ~				 \\
\cline{2-8}
~             	& Run309	& ZMO-b				& M3		& 1 	& 1 	& 20 			& 2966 		\\
\cline{2-8}
~             	& Run310	& ZMO-b				& Lum12	& 1 	& 1 	& 19--20 	& 2090 	 \\
~             	& ~				& enrLMO-t		& Lum11	& ~ 	& ~		& ~				& ~				 \\
\hline
CUPID R\&D			& ~				& LMO-1				& B297	& 2 	& 2 	& 15 			& 328 		 \\
\cline{3-8}
~								& ~				& LMO-2				& GeB		& 1 	& 1 	& 19 			& 201 		 \\
~								& ~				& LMO-3				& B304	& ~		& ~		& ~				& ~				 \\
\cline{3-8}
~								& ~				& enrZMO-t		& GeB		& 1 	& 1 	& 15 			& 717 		 \\
~								& ~				& enrZMO-b		& GeT		& ~		& ~		& ~				& ~				 \\
\cline{3-8}
~								& ~				& enrLMO-b		& GeOld	& 2 	& 4 	& 12 			& 487 		 \\
\hline
\end{tabular}
\end{center}
\label{tab:test}
%\end{table}
\end{table*}
\normalsize

An important difference between the set-ups is in the calibration procedure and the related policy. The CUPID R\&D is well suited for a regular control of the detector's energy scale in a wide energy range up to 2.6~MeV. On the contrary, a periodical calibration with the EDELWEISS-III set-up is available only with a $^{133}$Ba source ($\gamma$'s with energies below 0.4~MeV), while the insertion of a $^{232}$Th $\gamma$ source, as well as a few other available sources, requires lead/polyethylene shield opening, which is not supposed to be done frequently. Also, there is prohibition of the in-situ use of $^{55}$Fe sources for the light detectors calibration, which is not the case for the CUPID R\&D case. Finally, the control of the detector thermal response by a pulser system connected to the heaters is available for both set-ups.

%-------------------------------------------------------------------------------------------------
\subsection{Low-background measurements and data analysis}
\label{sec:Measurements}

The list of the low-background bolometric experiments and the main technical details are given in Table \ref{tab:test}. The bias currents of the order of a few nA were set to maximize the signal-to-noise ratio resulting in a working-point thermistor resistance of a few M$\Omega$. 

An optimum filter technique \cite{Gatti:1986,Radeka:1967} was used to evaluate the pulse height and shape parameters. It relies on the knowledge of the signal template and noise power spectrum; both are extracted from the data by averaging about 2 MeV energy signals (40 individual pulses) and baseline waveforms (5000 samples), respectively. For those detectors that were equipped with two temperature sensors, the data of the thermistor with the best signal-to-noise ratio were analyzed. The light-detector signal amplitude is estimated at a fixed time delay with respect to the heat signals as described in \cite{Piperno:2011}. Due to spontaneous temperature drifts, the amplitudes of the filtered signals from the crystal scintillator are corrected for the shift in thermal gain by using the heater pulses\footnote{The data of Run308 have been stabilized by using $\alpha$ events of $^{210}$Po from crystal contamination and it gives results similar to those of the heater-based stabilization method \cite{Armengaud:2015}.}. 

The heat response of the scintillating bolometers is calibrated with $\gamma$ quanta of $^{232}$Th (238.6, 338.3, 510.8, 583.2, 911.2, and 2614.5 keV), $^{40}$K (1460.8 keV) and/or $^{133}$Ba (356.0 keV) sources. The light detectors in the CUPID R\&D set-up were calibrated with the $^{55}$Mn X-ray doublet (5.9 and 6.5~keV) of the $^{55}$Fe source.

%-------------------------------------------------------------------------------------------------
\section{Performance of $^{100}$Mo-containing scintillating bolometers}
\label{sec:Performance}

\subsection{Time profile of the pulses}
\label{sec:Pulse}

The rising edge of the bolometric signal depends on the sensitivity of the sensor to athermal and/or thermal phonons created by a particle interaction and has a characteristic time ranging from microseconds (dominant athermal component) to milliseconds (dominant thermal component). Since the NTD sensors are sensitive mainly to thermal phonons, the rise time of the tested detectors given in Table \ref{tab:performance} is within the expectation. The heat detectors have longer leading edge (tens of ms) than that of light detectors (few ms) due to the larger volume and therefore to the larger heat capacity of the absorber.
 
The decaying edge time constant of the bolometric signal represents the thermal relaxation time, which is defined by the ratio of the heat capacity of the absorber to the thermal conductance to the heat bath. Therefore, it strongly depends on the material, the detector coupling to the heat bath and on the temperature. As one can see in Table \ref{tab:performance}, the variation of the decay time is even larger than that of the rise time, but again it has the typical values normally observed in light detectors (tens of ms) and massive bolometers (hundreds of ms). The improved coupling of the NTD sensor to the heat bath of B297 and B304 light detectors \cite{Coron:2004} leads to shorter decay time at the level of a few ms.

The only exception in the signal time constants of massive detectors is evident for both Zn$^{100}$MoO$_4$  crystals, the largest of all tested samples, which exhibit signals faster by about a factor 2 than the other devices, in particular those tested in the same set-up and at similar temperatures. It is interesting also to note that the bottom crystal is twice faster than the top one, as it was observed also in the test of the 60~g Zn$^{100}$MoO$_4$ detectors \cite{Barabash:2014}. The fast response of the enriched Zn$^{100}$MoO$_4$ bolometers has no clear explanation, but it is probably related to crystal quality. However, a fast detector response is crucial for a separation of the $^{100}$Mo $2\nu 2\beta$ events pile-ups \cite{Chernyak:2012,Chernyak:2014,Chernyak:2017}. Thus, this feature of Zn$^{100}$MoO$_4$ scintillators (e.g. $\tau_R$ below 10 ms) can lead to a better capability to discriminate random coincidences by heat pulse-shape analysis than that considered in Ref. \cite{Chernyak:2014}.

%-------------------------------------------------------------------------------------------
\subsection{Voltage sensitivity}
\label{sec:Sensitivity}

In a bolometric detector, the response to a nuclear event is a temperature rise directly proportional to the deposited energy and inversely proportional to the detector heat capacity. A thermistor converts the temperature variations to a voltage output, digitized by the readout system. Therefore, a bolometric response is characterized by a voltage sensitivity per unit of the deposited energy. 

A signal pulse height of the order of few tens to hundreds nV/keV is typical for NTD-instrumented massive bolometers. This figure corresponds to what is observed in all the tested crystals (see Table~\ref{tab:performance}). 

The reduced size of both the absorber and the sensor of light detectors (see Tables~\ref{tab:Bolo} and \ref{tab:LD}) leads to lower heat capacities and therefore to higher sensitivites, which are in the range $\sim$1--2~$\mu$V/keV for a good-performance 

%-------------------------------------------------------------------------------------------

\newpage
\clearpage

\begin{landscape}

\begin{table*}[htb]
\caption{Characteristics of $^{100}$Mo-containing scintillating bolometers. The pulse-shape time constants are the rise ($\tau_{R}$) and decay ($\tau_{D}$) times defined as the time difference between the 10\% and 
the 90\% of the maximum amplitude on the leading edge and the time difference between the 90\% and the 30\% of the maximum amplitude on the trailing edge, respectively. The signal sensitivity is measured as the thermistor voltage change for a unitary energy deposition. The intrinsic energy resolution (FWHM baseline) is determined by noise fluctuations at the optimum filter output. The energy resolution (FWHM) of light detectors was measured with a $^{55}$Fe X-ray source. The FWHM resolution of heat channels is obtained for $\gamma$ quanta of $^{40}$K, $^{133}$Ba, and $^{232}$Th $\gamma$ sources. $LY_{\alpha}$ and $LY_{\gamma(\beta)}$ denote light yields for $\alpha$s and $\gamma(\beta)$s, respectively. The quenching factor for $\alpha$ particles $QF_{\alpha}$ and the discrimination power $DP_{\alpha/\gamma(\beta)}$ (above 2.5 MeV)	are calculated according to the formulas given in the text}
%\footnotesize
%\scriptsize
\tiny
\begin{center}
\begin{tabular}{|c|l|ll|ll|lll|ll|}
 \hline
\bf{Scintillating} & Scintillator  & \multicolumn{2}{c|}{ZnMoO$_4$}    & \multicolumn{2}{c|}{Zn$^{100}$MoO$_4$} & \multicolumn{3}{c|}{Li$_2$MoO$_4$} & \multicolumn{2}{c|}{Li$_2$$^{100}$MoO$_4$} \\
\bf{bolometer}     & Crystal ID      	& \multicolumn{2}{c|}{ZMO-b}       & enrZMO-t   & enrZMO-b   & LMO-1      & LMO-2      & LMO-3      & enrLMO-t  & enrLMO-b\\
~                  & Size ($\oslash$$\times$h mm) & \multicolumn{2}{c|}{50$\times$40} & 60$\times$40 & 60$\times$40 & 40$\times$40 & 50$\times$40 & 50$\times$40 & 44$\times$40 & 44$\times$44 \\
~                  & Mass (g)         & \multicolumn{2}{c|}{334}       & 379      & 382      & 151      & 241      & 242      & 186 		& 204\\
\cline{2-11}
~                  & Light detector ID 			& M3    	& Lum12    	  & GeB  			 & GeT  			& B297  		 & GeB  		& B304  		& Lum11	 	  & GeOld \\
~                  & Size ($\oslash \times$h mm) & 44$\times$0.17 & 44$\times$0.17& 45$\times$0.30 & 45$\times$0.30 & 40$\times$0.045 & 45$\times$0.30 & 25$\times$0.030 & 44$\times$0.17 & 45$\times$0.30\\
\hline
\bf{Test}    			 & Underground lab	& LSM	& LSM	 & LNGS  & LNGS  & LNGS  & LNGS  & LNGS  & LSM			 & LNGS \\
~                  & $T_{base}$ (mK)  		& 18 & 19--20 & 15 & 15 & 15 & 19 & 19 & 19--20 & 12\\
\hline

\bf{Pulse-shape}	 & Light $\tau_R$ & 5.2	& 4.6		& 2.4 & 2.9 & 2.3 & 4.0	 & 2.7	& 3.5 & 5.1 \\
\bf{time constant} & Light $\tau_D$ & 23	& 24		& 12  & 14  & 2.6	& 8.5  & 6.4	& 13  & 13 \\
\bf{(ms)}               & Heat $\tau_R$	& 19	& 38		& 9.6 & 6.4 & 17  & 29   & 30		& 27  & 18 \\
~                  & Heat $\tau_D$	& 200	& 204		& 37  & 18  & 67  & 339  & 414	& 169 & 88 \\
 \hline
\bf{Sensitivity} 	 & Light detector & 2200$^*$ & 2500$^*$ & 1047 & 1053  & 4030 & 850	& 15800	& 1900$^*$ & 2910 \\
\bf{(nV/keV)}					 & Heat detector  & 48				 & 26					& 73 	 & 39		 & 166  & 11	& 23		& 32				 & 89 \\
 \hline
\bf{Light FWHM} 	& Baseline  			& $\sim$140$^*$ & $\sim$60$^*$ & $\sim$490  & $\sim$230  & $\sim$42 & $\sim$420  & $\sim$18	& $\sim$70$^*$ & $\sim$140 \\
\bf{(eV)}							& X-ray $^{55}$Mn, 5.9 keV  & --				 & -- 				& 787(3)  & 289(1)  & 334(4)  & 555(5)  & 504(4)	& --				 & 303(2) \\
 \hline
\bf{Heat FWHM} 		& Baseline   					& $\sim$1.6 	 & $\sim$3.6  	& $\sim$2.6	  & $\sim$4.3		& $\sim$0.6			& $\sim$2.3  		& $\sim$1.6  		& $\sim$1.2		 	& $\sim$1.2 \\
\bf{(keV})							& $\gamma$ $^{133}$Ba, 356 keV   & 3.7(1) & 5.1(1) & --	  	& --    & --  		& --  		& --  		& 2.54(4)	& -- \\
~                	& $\gamma$ $^{228}$Ac, 911 keV   & 4.4(7) & 10(1)  & 5.6(7)  & 9(2)  & 2.0(3)  & 3.9(6)  & 3.1(6) 	& 3.1(5)	& 3.1(2) \\
~                	& $\gamma$ $^{40}$K, 1461 keV  & -- 		 & 7.9(2) & 6.7(6)  & 14(1) & -- 	  	& 4.2(3) 	& 4.4(3) 	& 4.1(2) 	& -- \\
~                	& $\gamma$ $^{208}$Tl, 2615 keV  & 9(1)   & 12(1)  & 9.1(7)  & 22(2)	& 3.8(6)  & 6(1)    & 4.7(7)  & 6.3(6)	& 5.0(5) \\
~                	& $\alpha$ $^{210}$Po, 5407 keV  & 8.8(1) & 9.0(2) & $\sim$47 & $\sim$100 & 7(2) & 9(1)	& 9(2)  	& 5.4(3) 	& --  \\
 \hline
\bf{Response to}  & $LY_{\gamma(\beta)}$ (keV/MeV) 					  	& -- 	 & --   & 1.32(1)		& 1.20(1)  & 0.68(4)  & 0.99(1)  & 0.121(2)$^{**}$ & --		 & 0.775(4) \\
\bf{$\gamma(\beta)$ and $\alpha$}	& $LY_{\alpha}$ (keV/MeV)			& -- 	 & --   & 0.217(4)	& 0.148(2)  & 0.165(1)  & 0.203(4)  & 0.0236(3) & -- 	 & 0.153(2) \\
~                	& $QF_{\alpha}$																& 0.15 & 0.17 & 0.17	& 0.13  & 0.23  & 0.20  & 0.17 & 0.22  & 0.19 \\
~                	& $DP_{\alpha/\gamma(\beta)}$ & 12 	 & 21 	& 7.8 	& 11 		& 16 	& 8.7 	& 11 	 & 18	 	 & 12 \\
 \hline
\multicolumn{11}{l}{* --- Estimations are based on rough calibrations by scintillation light (see Section \ref{sec:Sensitivity}).} \\ 
\multicolumn{11}{l}{** --- Low light yield is caused by non-optimal light collection conditions of the measurements (see Section \ref{sec:PSDly}).} \\ 
\end{tabular}
\end{center}
\label{tab:performance}
%\end{table}
\end{table*}
\normalsize

\end{landscape}

%\clearpage
\newpage

%-------------------------------------------------------------------------------------------------

\noindent detector. This is the case for all the tested light 
detectors\footnote{In order to estimate the performance of the light detectors operated in the EDELWEISS-III set-up, we roughly calibrated them by using the heat-light data of $^{232}$Th runs and assuming that the light yield (see Sec. \ref{sec:PSDly}) for $\gamma(\beta)$ events is equal to 1.0 and 0.77~keV/MeV for ZnMoO$_4$ and Li$_2$$^{100}$MoO$_4$ detectors, respectively. The assumption about the scintillation yield of ZnMoO$_4$ is based on early investigations of the similar size detectors. In the case of Li$_2$$^{100}$MoO$_4$, we expect similar scintillation properties of the samples produced from the same boule.} (see in Table \ref{tab:performance}), except the ones with even smaller size (B297 and B304) and subsequently sensitivity enhanced by up to one order of magnitude.

%-------------------------------------------------------------------------------------------------
\subsection{Energy resolution}
\label{sec:FWHM}

Most of the used light detectors have similarly good performance also in terms of energy resolution, in particular their baseline noise is $\sim$0.14--0.5~keV FWHM (see in Table \ref{tab:performance}). 
The only exceptions are detectors with enhanced sensitivity (B297 and B304) for which the baseline noise is below $\sim$0.05 keV FWHM. However, they also exhibit a strong position-dependent response, therefore the energy resolution measured with an uncollimated $^{55}$Fe source is near to that obtained with the other light detectors (FWHM $\sim$ 0.3--0.8~keV at 5.9~keV).

As it was mentioned above, better than 10~keV FWHM energy resolution at the ROI is one of the most crucial requirements for cryogenic double-beta decay detectors. This goal was successfully achieved with both natural and $^{100}$Mo-enriched ZnMoO$_4$ and Li$_2$MoO$_4$ based bolometers\footnote{The operation of the ZMO-t bolometer, a twin of the ZMO-b, was severely affected by an insufficient tightening of the PTFE elements, therefore we omit quoting its performance.} (see Table \ref{tab:performance}). Below we discuss the obtained results.

\begin{figure*}[htbp]
  \includegraphics[width=0.48\textwidth]{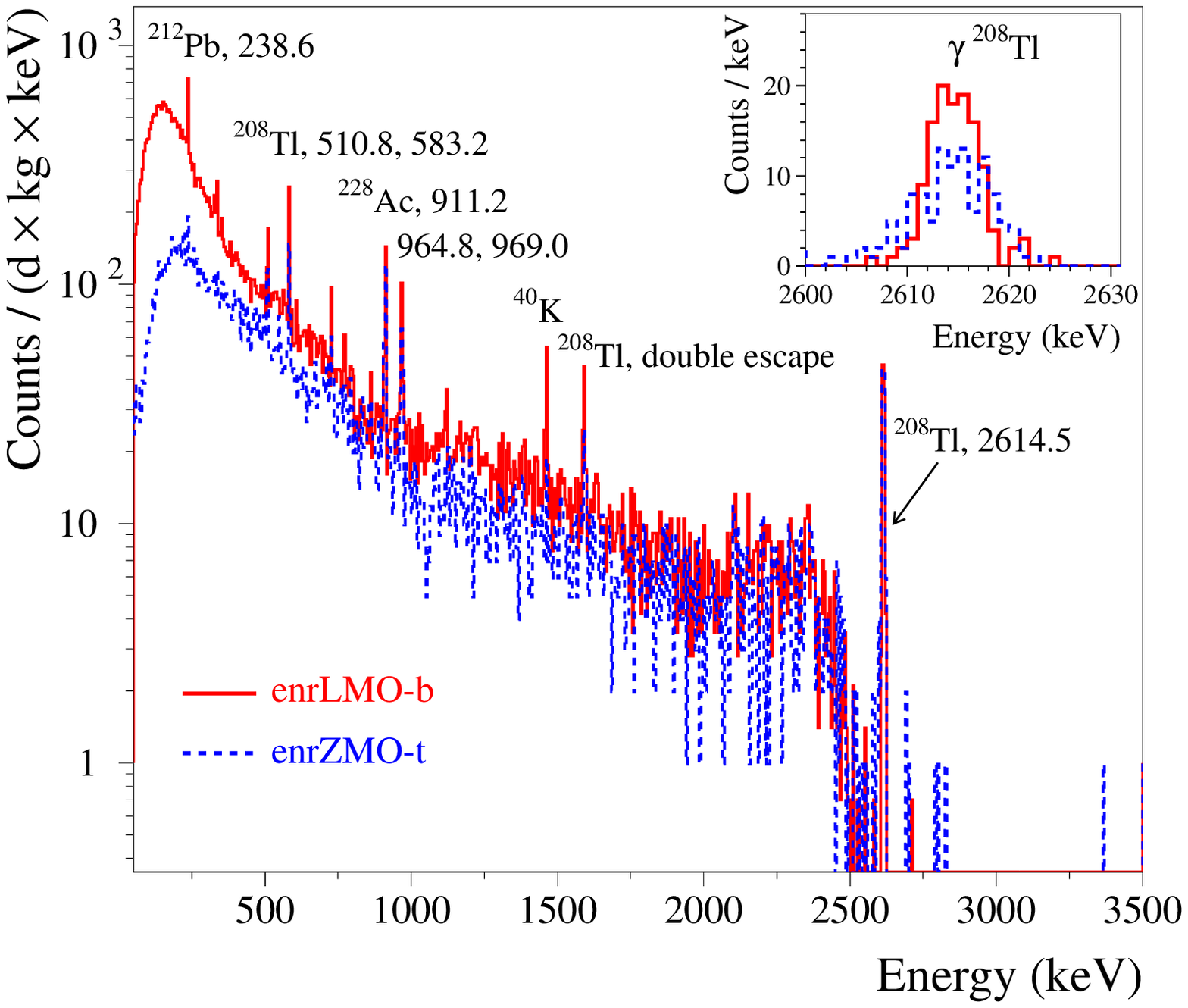}
  \includegraphics[width=0.48\textwidth]{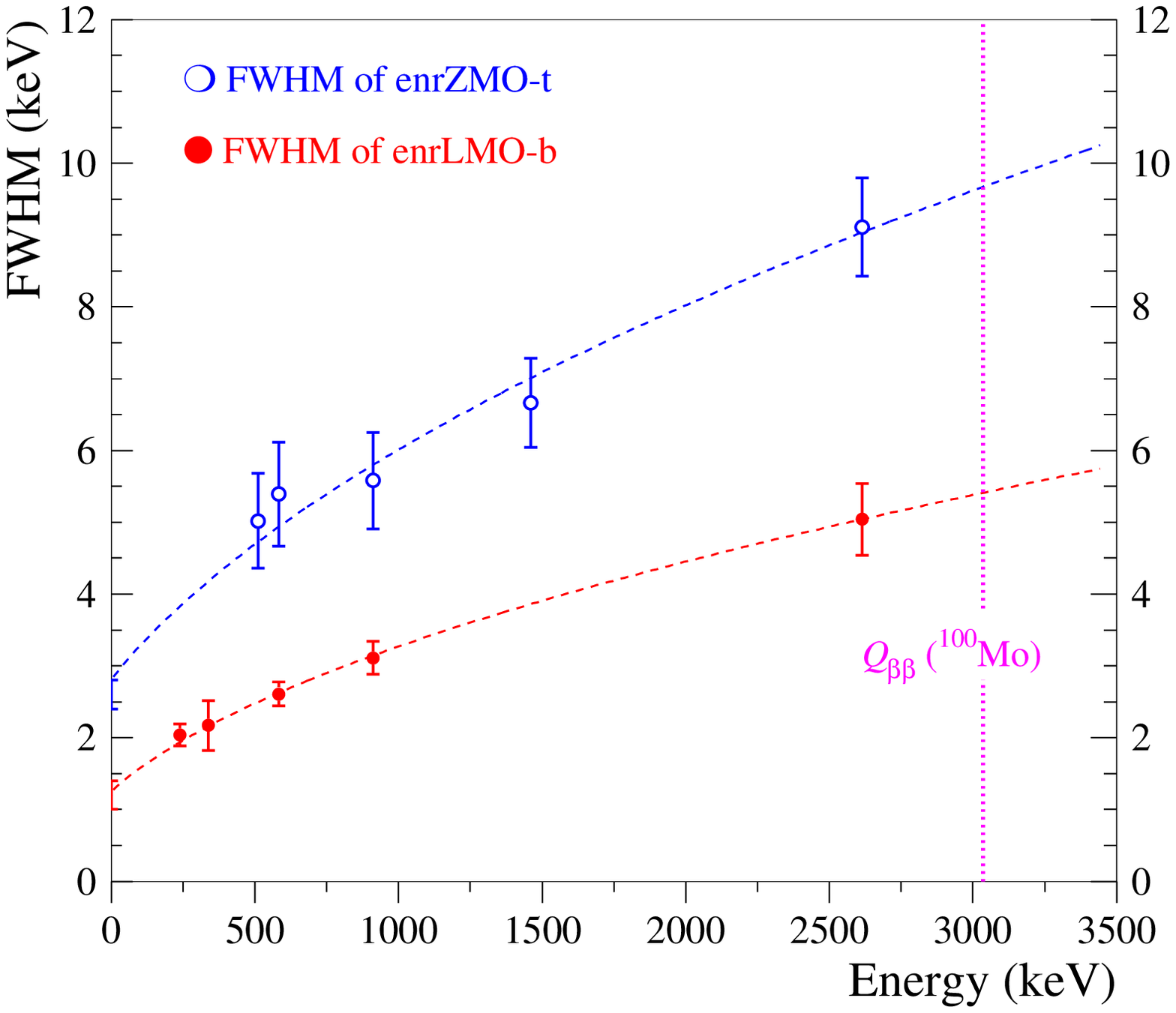}
\caption{The energy spectra of the $^{232}$Th $\gamma$ source measured by the $\sim$0.4~kg Zn$^{100}$MoO$_4$ (enrZMO-t; dashed histogram) and 0.2~kg Li$_2$$^{100}$MoO$_4$ (enrLMO-b; solid histogram) bolometers over 64 h and 168 h, respectively, at LNGS (left figure). The energy bin is 5~keV. The 2615~keV peak of the $^{208}$Tl $\gamma$ quanta accumulated by the detectors is shown in the inset.  
The energy dependence of the energy resolution of the ZMO-t and enrLMO-b detectors (right figure). 
The fits to the data by a function FWHM (keV) = $\sqrt(p_1 + p_2 \times E_{\gamma}(keV))$ ($p_1$ and $p_2$ are free parameters) are shown by the dashed lines. The parameters of fits are 7.9(5)~keV$^2$ and 0.0282(6)~keV for the ZMO-t and 1.6(2)~keV$^2$ and 0.0091(3)~keV for the enrLMO-b. The dotted line indicates the $Q_{\beta\beta}$ value of $^{100}$Mo (3034 keV)}
\label{fig:DETcali}
\end{figure*} 

The Li$_2$MoO$_4$ detectors exhibit twice better energy resolution than the ZnMoO$_4$ ones and the achieved values of 4--6~keV FWHM at 2615~keV are at the level of the best resolutions ever obtained with massive bolometers \cite{Artusa:2014a,Artusa:2014b,Alduino:2016}. In particular, the energy resolution of Li$_2$MoO$_4$ bolometers is comparable to the performance of the TeO$_2$ cryogenic detectors of the CUORE-0 experiment (the effective mean FWHM at 2615~keV is 4.9~keV with a corresponding RMS of 2.9~keV \cite{Alduino:2016}). This is mainly due to the fact that Li$_2$MoO$_4$, as TeO$_2$, demonstrates a low thermalization noise, i.e. a small deviation of the energy resolution from the baseline noise width. The results of the ZnMoO$_4$ and Li$_2$$^{100}$MoO$_4$ detectors show possible improvement of the energy resolution by lowering of the temperature, as it is expected thanks to increased signal sensitivity. A dependence of the performance on the sample position in the Zn$^{100}$MoO$_4$ boule, observed early with small \cite{Barabash:2014} and now with large samples, is also evident. It could be related to the degradation of the crystal quality along the boule. Thanks to the higher crystal quality, no such effect is observed for Li$_2$$^{100}$MoO$_4$ crystals.

The energy spectra of a $^{232}$Th $\gamma$ source measured by the $^{100}$Mo-enriched bolometers (enrZMO-t and enrLMO-b)  and the corresponding energy-dependance of the heat-channel resolution are illustrated in Fig. \ref{fig:DETcali}. The chosen data of Zn$^{100}$MoO$_4$ and Li$_2$$^{100}$MoO$_4$ detectors represent the typical energy resolution for bolometers based on these materials in case of optimal experimental conditions (Table \ref{tab:performance}). Using the fitting parameters for the curves shown in Fig. \ref{fig:DETcali} (right), the expected energy resolution of the enrZMO-t and enrLMO-b cryogenic detectors at $Q_{\beta\beta}$ of $^{100}$Mo is 9.7$\pm$0.1~keV and 5.4$\pm$0.1~keV, respectively. Thus, the energy resolution of the Zn$^{100}$MoO$_4$ detectors is acceptable but needs still an optimization, while Li$_2$$^{100}$MoO$_4$ bolometers already meet the resolution required for future generation bolometric $0\nu2\beta$ experiments \cite{Beeman:2012,Beeman:2012a,Artusa:2014a,CUPID}.

%-------------------------------------------------------------------------------------------------
\subsection{Response to $\alpha$s and particle identification capability}
\label{sec:PSDly}

\subsubsection{Scintillation-assisted particle discrimination}

\begin{figure*}[htbp]
\centering
  \includegraphics[width=0.48\textwidth]{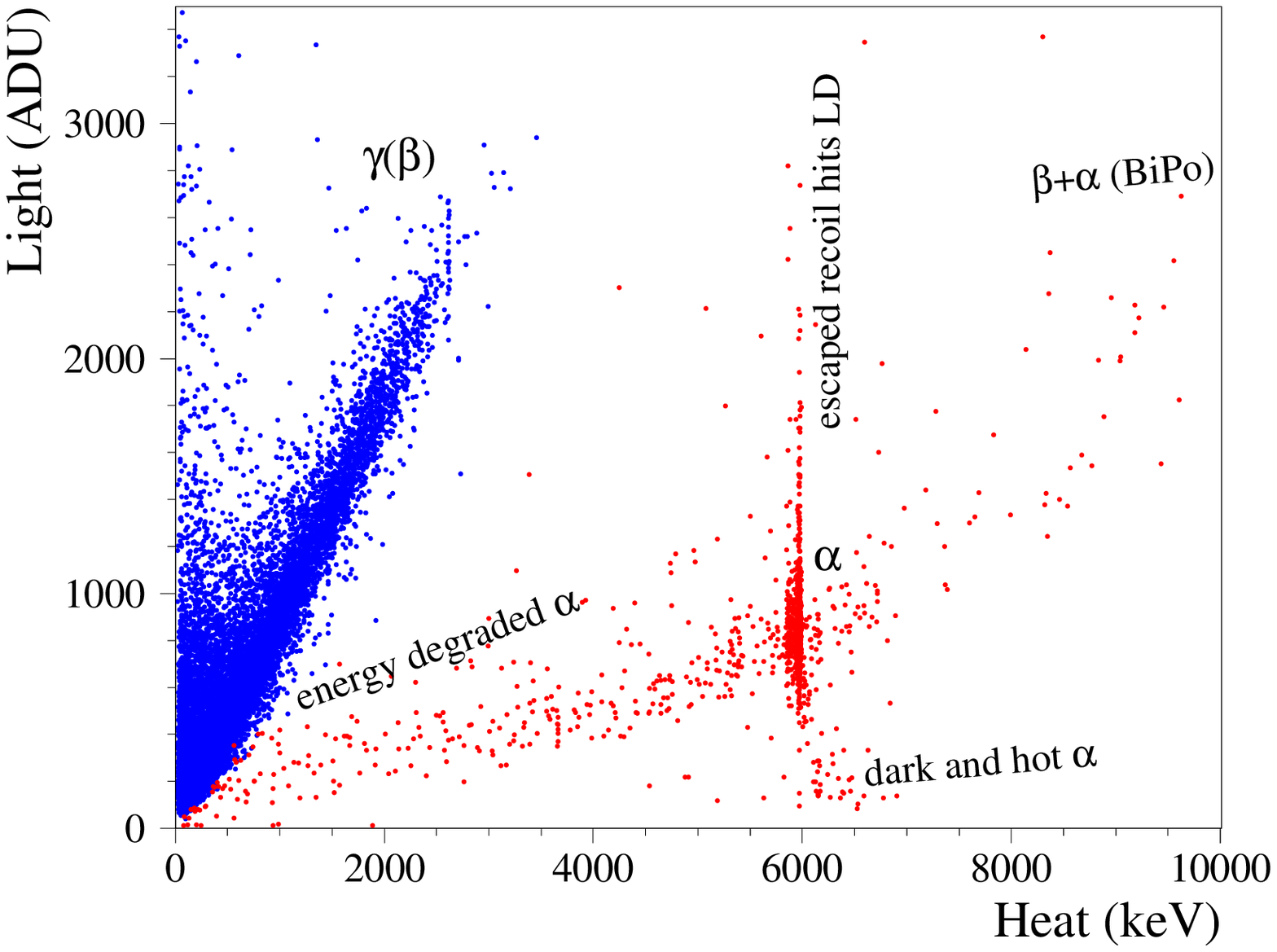}
  \includegraphics[width=0.48\textwidth]{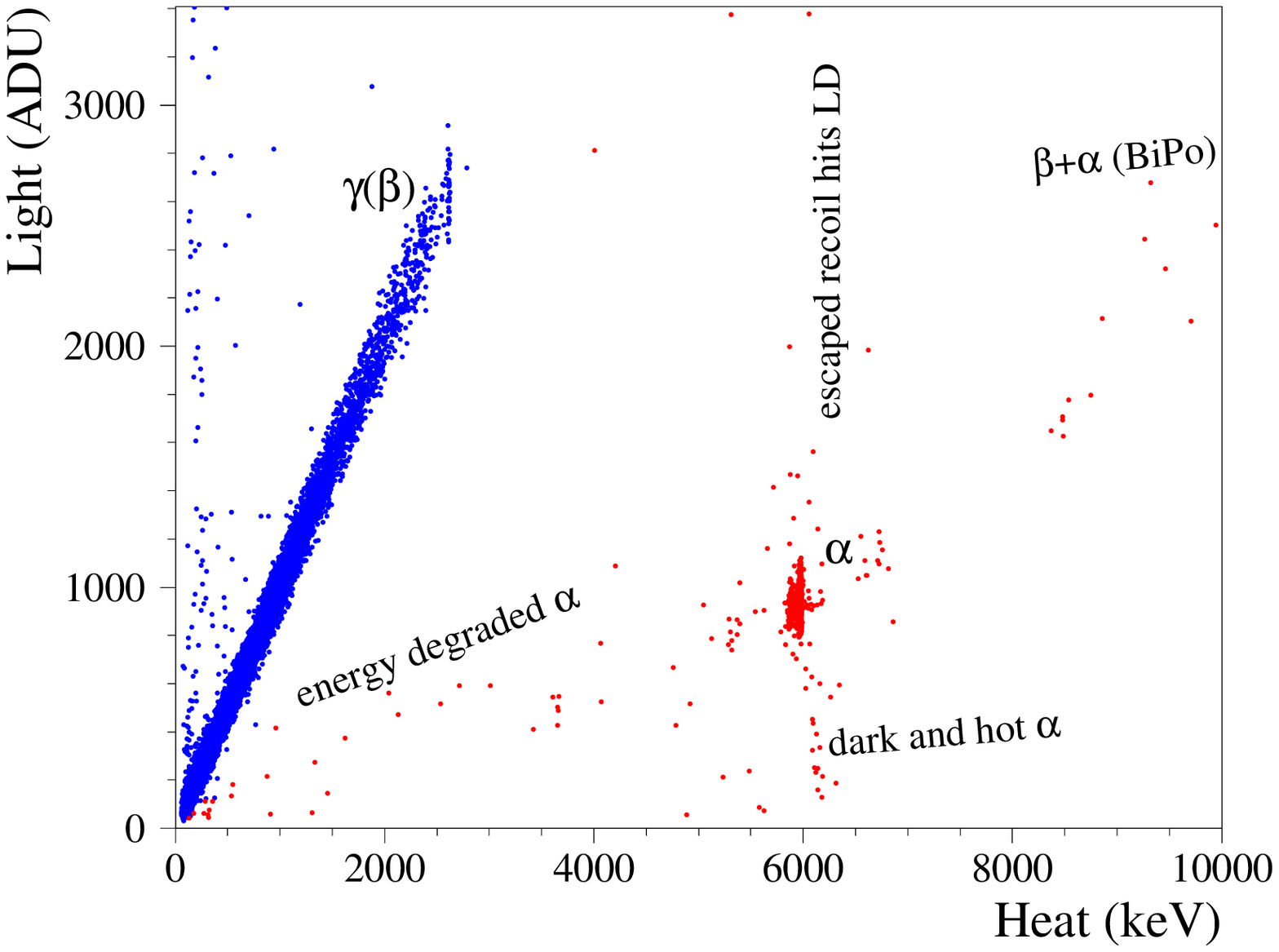}
  \includegraphics[width=0.48\textwidth]{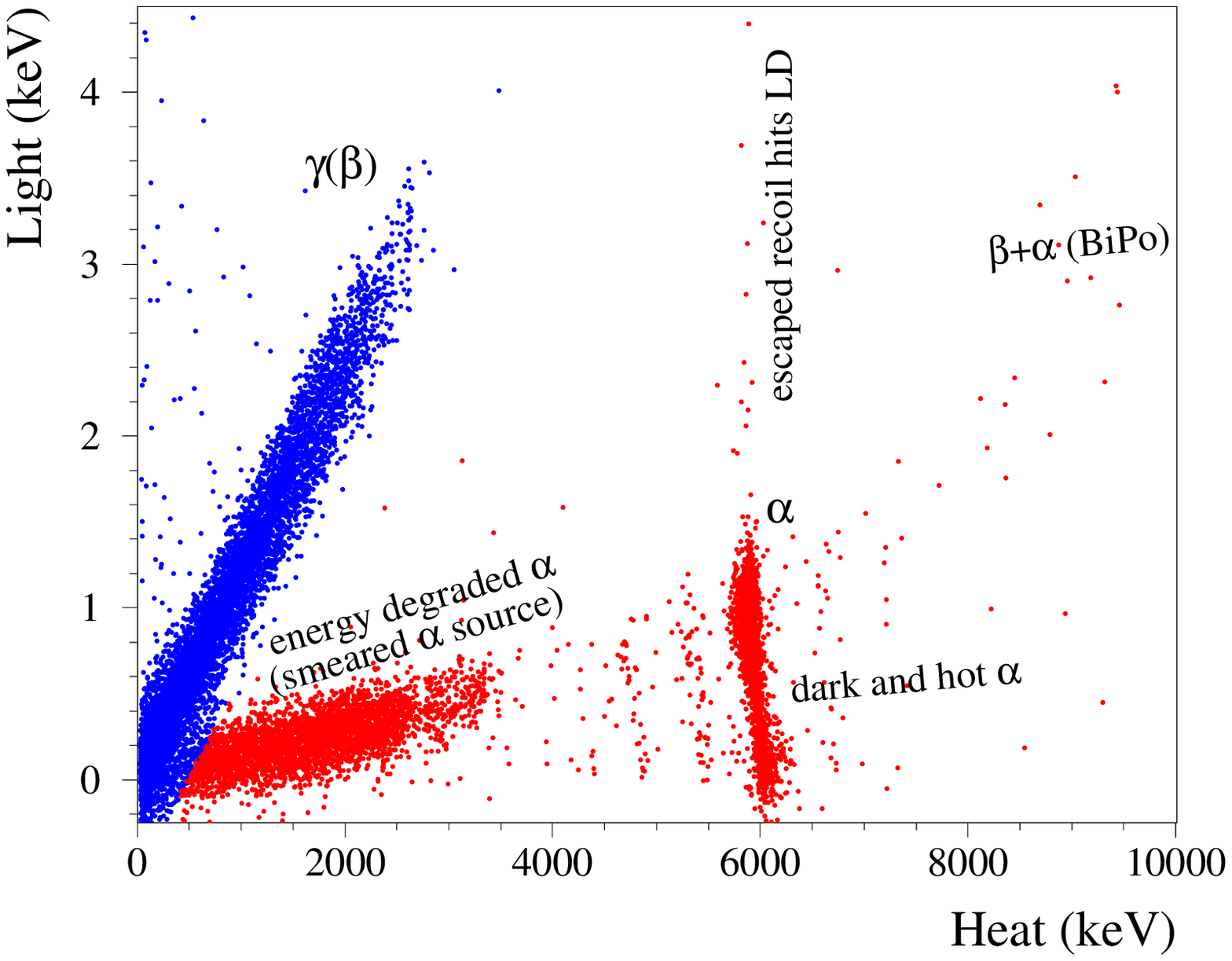}
  \includegraphics[width=0.48\textwidth]{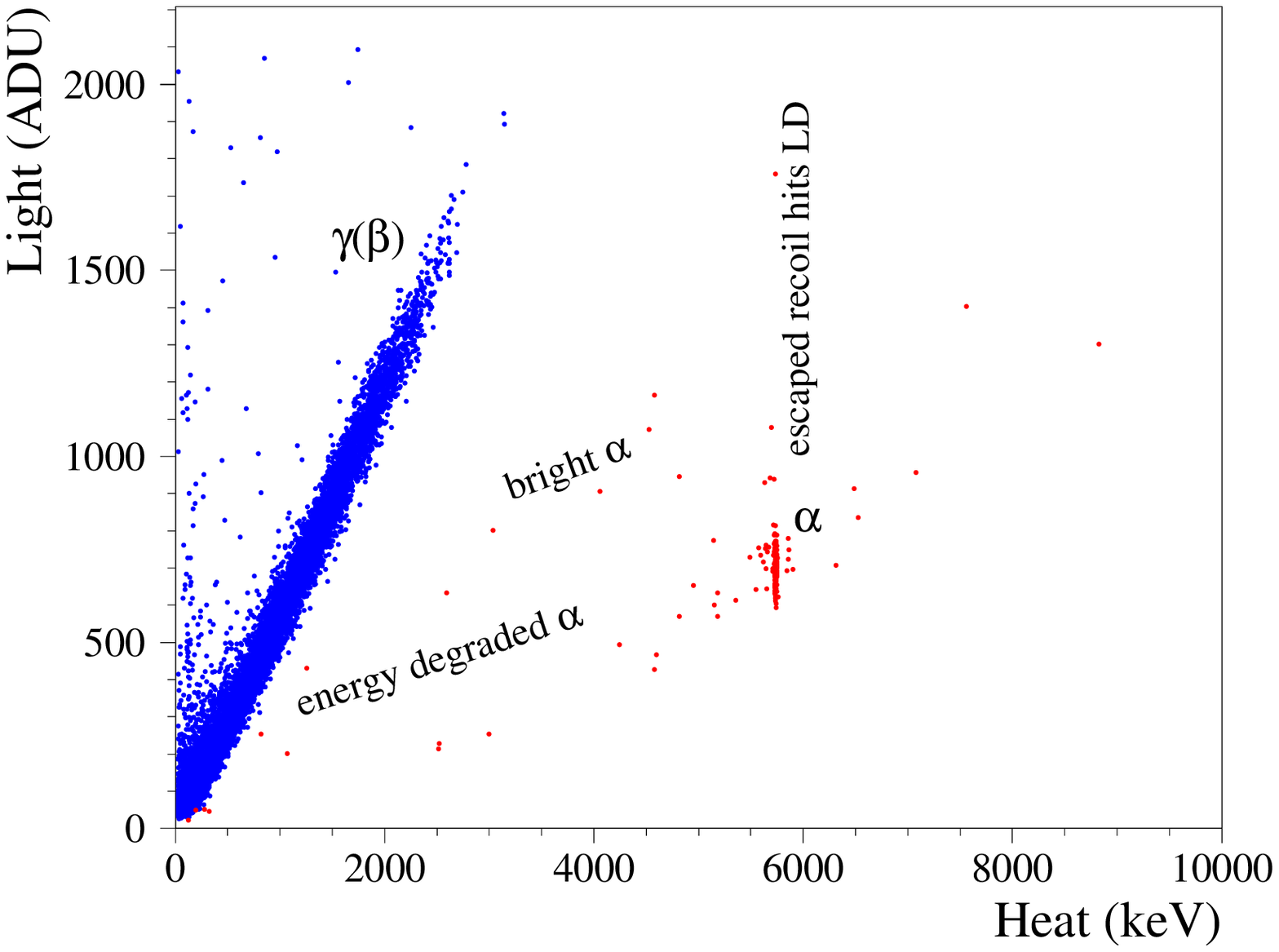}
\caption{Scatter plots of light-versus-heat signals of the background data collected with scintillating bolometers based on 334~g ZnMoO$_4$ (ZMO-b, top figures), 382~g Zn$^{100}$MoO$_4$ (enrZMO-b, bottom left), and 186~g Li$_2$$^{100}$MoO$_4$ (enrLMO-t, bottom right) crystals over 2767, 1300, 593, and 1303~h, respectively. The Zn$^{100}$MoO$_4$ detector was operated in the CUPID R\&D cryostat, while the other data were accumulated in the EDELWEISS-III set-up (the light signals of the latter are in analog-to-digit units, ADU). The heat channels were calibrated with $\gamma$ quanta. The $\gamma$($\beta$) and $\alpha$ events populations are distinguished in color by using the cuts on the heat energy and the light yield parameters (see the text). The particle identification capability of the ZnMoO$_4$ detector affected by vibration noise (top left) was substantially improved in the suspended tower (top right).  The features of the $\alpha$ particle populations are discussed in the text}
\label{fig:light_vs_heat}
\end{figure*}

By using coincidences between the heat and the light channels, one can plot a light-vs-heat scatter plot as the ones presented in Fig. \ref{fig:light_vs_heat}. The heat channel of all data shown in Fig. \ref{fig:light_vs_heat} is calibrated by means of $\gamma$ quanta of the calibration sources and it leads to $\sim$10\% heat miscalibration for $\alpha$ particles due to a so-called thermal quenching, common for scintillating bolometers (e.g. see results for different scintillators in \cite{Beeman:2012,Cardani:2013,Artusa:2016,Arnaboldi:2010}). Therefore, in order to present the correct energy of the $\alpha$ events, an additional calibration based on the $\alpha$ peaks identification is needed.

\nopagebreak

As it is seen in Fig. \ref{fig:light_vs_heat}, the light-vs-heat scatter plot contains two separated populations: a band of $\gamma$($\beta$)'s and a distribution of events associated to 
$\alpha$ decays. This is due to the fact that the amount of light emitted in an oxide scintillator by $\alpha$ particles is quenched typically to $\approx$ 20\% with respect to $\gamma$ quanta 
($\beta$ particles) of the same energy (see, e.g., Ref. \cite{Tretyak:2010}). Therefore, the commonly used particle identification parameter for scintillating bolometers is the light yield ($LY$), that we will define as a ratio of the light-signal amplitude measured in keV to the heat-signal amplitude measured in MeV. 

The data of all detectors with directly calibrated light channel (all measurements at LNGS) have been used to determine the $LY_{\gamma(\beta)}$ and $LY_{\alpha}$ values of $\gamma$($\beta$) and $\alpha$ events selected in the heat-energy range 2.5--2.7~MeV and 2.5--7~MeV\footnote{The interval for the selection of $\alpha$s is reduced to 2.5--3.5~MeV for the measurements performed with a smeared $\alpha$ source (enrZMO-t, enrZMO-b, and LMO-2 detectors).}, respectively. In spite of the quite evident constancy of the $LY_{\gamma(\beta)}$ in a wide energy range (as it is seen from the slop of $\gamma(\beta)$s in Fig. \ref{fig:light_vs_heat}), the event selection was applied above 2.5 MeV, because the same distributions have been used to calculate $\alpha$/$\gamma(\beta)$  discrimination power (see below) close to the ROI of $^{100}$Mo. The $LY$ values extracted from the present data are given in Table \ref{tab:performance}. 

The light yields for $\gamma(\beta)$ events measured with both Zn$^{100}$MoO$_4$ scintillating bolometers are in the range 1.2--1.3~keV/MeV, similar to the results of previous investigation of natural (see \cite{Beeman:2012b,Berge:2014} and references therein) and $^{100}$Mo-enriched \cite{Barabash:2014} ZnMoO$_4$ detectors. Thanks to the progress in the development of high quality lithium molybdate scintillators --- as documented in the present work and recently in Ref. \cite{Velazquez:2017} --- the $LY_{\gamma(\beta)}$ values for Li$_2$MoO$_4$ and Li$_2$$^{100}$MoO$_4$ scintillation bolometers, which lay in the range 0.7--1~keV/MeV, become comparable to the light yields of the ZnMoO$_4$ detectors. The improvement of the $LY$ with respect to the early investigations with Li$_2$MoO$_4$ detectors \cite{Barinova:2010,Cardani:2013} is of about a factor of 2. One Li$_2$MoO$_4$ bolometer (LMO-3) was viewed by a light detector with a significantly lower area implying a reduced light collection and consequently a rather small $LY_{\gamma(\beta)}$ value of 0.12 keV/MeV. 

The ratio of the $LY$ parameters for $\alpha$s and $\gamma$($\beta$)s gives the quenching factor of the scintillation light signals for $\alpha$ particles: $QF_{\alpha} = LY_{\alpha}/LY_{\gamma(\beta)}$. An absolute light detector calibration is not needed to calculate this parameter. As it is seen in Table \ref{tab:performance}, the results for ZnMoO$_4$ and Li$_2$MoO$_4$ detectors are similar showing $\approx$ 20\% quenching of the light emitted by $\alpha$ particles with respect to the $\gamma$($\beta$) induced scintillation.

The efficiency of discrimination between $\alpha$ and $\gamma(\beta)$ populations can be characterized by the so-called discrimination power $DP_{\alpha/\gamma(\beta)}$ parameter defined as: 

\begin{center}
$DP_{\alpha/\gamma(\beta)} = \left|\mu_{\gamma(\beta)}-\mu_{\alpha}\right|/\sqrt{\sigma_{\gamma(\beta)}^2+\sigma_\alpha^2}$, 
\end{center}

\noindent where $\mu$ ($\sigma$) denotes the average value (width) of the $\alpha$ or $\gamma$($\beta$) distribution. The $DP_{\alpha/\gamma(\beta)}$ value is estimated for $\gamma$($\beta$) and $\alpha$ events selected for the $LY$ determination (see above). 

\begin{figure}[htbp]
\centering
  \includegraphics[width=0.48\textwidth]{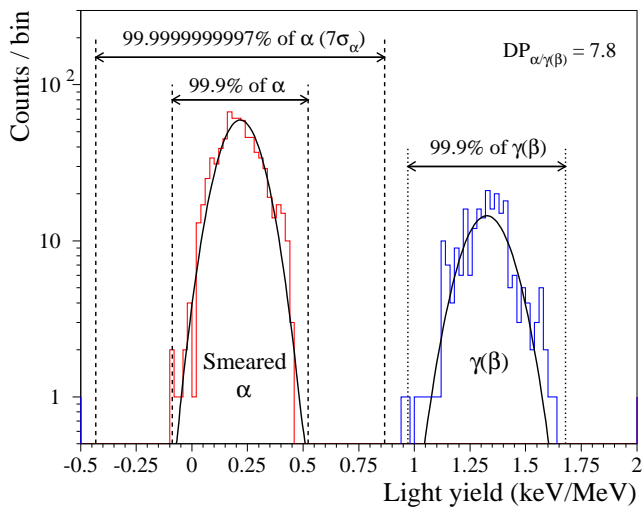}
\caption{The light yield distributions of $\alpha$ particles from a smeared $\alpha$ source and $\gamma$($\beta$) events collected by a 379 g Zn$^{100}$MoO$_4$ scintillating bolometer (enrZMO-t) over 593 h of background and 78 h of $^{232}$Th calibration measurements in the CUPID R\&D set-up at LNGS (Italy). The distributions are fitted by Gaussian functions shown by solid lines. The corresponding discrimination power is $DP_{\alpha/\gamma(\beta)}$ = 7.8. The intervals containing 99.9\% of both event types and $\pm$7 sigma interval of the $\alpha$ band are also given}
\label{fig:DP}
\end{figure}

As reported in Table \ref{tab:performance}, the achieved discrimination power for all the tested detectors is $DP_{\alpha/\gamma(\beta)}$ = 8--21, which implies a high level of the $\alpha$/$\gamma$($\beta$) separation: more than 99.9\% $\alpha$ rejection while preserving practically 100\% $0\nu 2\beta$ signal selection efficiency. The separation efficiency is illustrated in Fig. \ref{fig:DP} for the scintillating bolometer enrZMO-t with the lowest achieved $DP_{\alpha/\gamma(\beta)}$ due to the modest performance of the GeB light detector. It is to emphasize the $LY_{\gamma(\beta)}$ $\sim$ 0.1 keV/MeV obtained with the LMO-3 detector, which would not allow effective  particle identification by using a standard-performance light detector with 0.2--0.5 keV FWHM baseline noise\footnote{This is the case for Cherenkov light tagging in TeO$_2$ bolometers; e.g. see Ref. \cite{Artusa:2017}.}. However, the performance of the B304 optical bolometer --- which featured 0.02 keV FWHM baseline noise --- was high enough to provide highly-efficient particle identification even with this detector ($DP_{\alpha/\gamma(\beta)}$ = 11). 

\nopagebreak

\subsubsection{Peculiarities in particle identification}

Fig. \ref{fig:light_vs_heat} illustrates observed peculiarities of some detectors which could affect the particle identification capability. These peculiarities are originated either by a noise-affected detector performance or by a feature of the detector's response to $\alpha$s, which exhibits classes of events with more quenched or enhanced light signals. Below we will discuss briefly these observations and their impact on background in a $0\nu 2\beta$ decay experiment with $^{100}$Mo.

High vibrational noise in a light detector affects the precision of the light-signal amplitude evaluation, especially for events with a low scintillation signal ($\alpha$ events and $\gamma$($\beta$)s below $\sim$1~MeV). This was an issue of the measurements with ZnMoO$_4$ detectors in Run308 and Run309, and this effect is apparent in Fig. \ref{fig:light_vs_heat} (top left). The problem can be solved by using a mechanically isolated system inside a cryostat (see Table~\ref{tab:setup} and e.g. Refs.~ \cite{Pirro:2006b,Lee:2017,Olivieri:2017}). In particular, a stable and reliable light-channel performance of the ZnMoO$_4$ scintillating bolometer in the suspended tower (Run310) is evident from Fig. \ref{fig:light_vs_heat} (top right).

The data of natural and $^{100}$Mo-enriched ZnMoO$_4$ bolometers contain some $\alpha$ events that have more quenched light output and enhanced heat signals; e.g. see ``dark hot $\alpha$'' in Fig. \ref{fig:light_vs_heat}. In the past, the same effect was observed in bolometric tests of small ZnMoO$_4$ \cite{Gironi:2010} and Zn$^{100}$MoO$_4$ \cite{Mancuso:2016} crystals which also exhibit defects and macro inclusions. A major part of such events is distributed close to $^{210}$Po ($\alpha$ structures at around 6 MeV in electron-equivalent energy in Fig. \ref{fig:light_vs_heat}), the main contamination of the investigated ZnMoO$_4$ crystals. Only a short-range ($\alpha$) interaction in the crystal bulk exhibits this anomaly, because it is not evident either for $\alpha$ interactions at the crystal surface (energy-degraded $\alpha$ events) or for $\gamma$($\beta$) events which have longer mean path in the crystal than the bulk $\alpha$'s. This phenomenon is probably related to the thermal quenching, as suggested by the pronounced anti-correlation between light and thermal signals in the $\alpha$ response. The effect is more evident for the enriched crystals, which contain more inclusions than the natural ones: e.g. about 40\% of $^{210}$Po events acquired by the enrZMO-b detector are attributed to the ``dark hot $\alpha$'', while four times lower amount of such events is observed in the ZMO-b bolometer. Thereby, the origin of this anomaly in the response to $\alpha$ interactions is probably related to the crystals imperfections. Taking into account that two electrons are expected in the $0\nu 2\beta$ of $^{100}$Mo, we expect that the $0\nu 2\beta$ signal is unaffected by this anomaly. Furthermore, it does not affect the detector's capability to identify and reject the $\alpha$-induced surface events, which constitute the most challenging background in a bolometric $0\nu 2\beta$ experiment without particle identification. Negative effects are only expected on the precision of the $\alpha$ spectroscopy, which however is important not only to build a background model through radiopurity determination, but also for the off-line rejection of $\alpha$-$\beta$ delayed events from decays of $^{212,214}$Bi-$^{208,210}$Tl \cite{Beeman:2012,Pirro:2006,Beeman:2012b}. 

The light-vs-heat data of several scintillating bolometers contain also $\alpha$ events with an enhanced light signal with respect to those of the prominent $\alpha$ distribution. As it is seen in Fig. \ref{fig:light_vs_heat}, these events belong to three families: BiPo events, surface events with escaped nuclear recoils hitting the light detector, and the so-called ``bright $\alpha$''. 

The first family consists of unresolved coincidences in $^{212,214}$Bi-$^{212,214}$Po decays (BiPo events in Fig. \ref{fig:light_vs_heat}). Due to the slow bolometric response, the $\beta$ decays of $^{212}$Bi ($Q_{\beta}$ = 2254~keV) and $^{212}$Bi ($Q_{\beta}$ = 3272~ keV) overlap with subsequent $\alpha$ decays of $^{212}$Po ($Q_{\alpha}$ = 8954~keV, $T_{1/2}$ = 0.3~$\mu$s) and 
$^{214}$Po ($Q_{\alpha}$ = 7833~keV, $T_{1/2}$ = 164~$\mu$s), respectively. Therefore, they are registered as a single event with a heat energy within 8--11 MeV range and a light signal higher than that of a pure $\alpha$ event of the same energy. Since the BiPo  events are distributed far away from 3 MeV, they have no impact on the $0\nu 2\beta$ ROI of $^{100}$Mo. 

In a case of an $\alpha$ decay on a crystal surface, a nuclear recoil (or an $\alpha$ particle) can escape from the scintillator and hit the light detector. Such events belong to the second family indicated in Fig. \ref{fig:light_vs_heat}. Taking into account that only a few keV energy-degraded recoil can mimic a light signal of ZnMoO$_4$ or Li$_2$MoO$_4$ bolometer, the heat energy release has to be close to the nominal $Q_{\alpha}$-value additionally enhanced due to the thermal quenching. Therefore, independently on the surface $\alpha$ activity of radionuclides from U/Th chains (4--9 MeV $Q_{\alpha}$-values), they cannot populate the ROI of $^{100}$Mo. Among other natural $\alpha$-active nuclides, a probable contaminant is $^{190}$Pt ($Q_{\alpha}$ = 3252~keV \cite{Wang:2017a}) due to the crystal growth in a platinum crucible. However, even in such case the expected heat signal is about 0.5 MeV away from the $Q_{\beta\beta}$ of $^{100}$Mo, as well as the $^{190}$Pt bulk contamination in the studied crystals is expected to be on the level of a few $\mu$Bq/kg \cite{Armengaud:2015}. We can therefore conclude that also this class of events does not play a role in the search for the $0\nu 2\beta$ decay of $^{100}$Mo. 

The last family --- consisting of ``bright $\alpha$'' events in Fig. \ref{fig:light_vs_heat} --- stem from the documented scintillation properties of the reflecting film. Specifically, an energy deposition in this film can take place for surface-originated $\alpha$ decays, which can produce a heat and a light signal in the scintillating crystal but also a flash of scintillation light from the reflecting film, which adds up to that of the crystal scintillator. This results into an enhanced light signal. Consequently, the population of energy-degraded $\alpha$ events can leak to the ROI of $^{100}$Mo in the heat-light scatter plot, prviding an unavoidable background. To check the scintillation response of the 3M film, we have performed a test using a photomultiplier and a $^{238}$Pu $\alpha$ source. The observed scintillation is at the level of 15\%--34\% relatively to NE102A plastic scintillator (depending on the side of the film facing the photomultiplier). Therefore, such a feature of the reflector spoils the particle discrimination capability of the detector. In order to solve this issue, a reflecting material without scintillation properties has to be utilized or the reflecting film has to be omitted\footnote{The results of the recently-completed Run311 in the EDELWEISS-III set-up, in which both Li$_2$$^{100}$MoO$_4$ detectors enrLMO-t and enrLMO-b were operated without the reflecting foil, demonstrate the capability of $\alpha$ particle discrimination at the level of 9 sigma in spite of half light-collection efficiency resulting in $\sim$0.4~keV/MeV light yield \cite{Poda:2017}.}.

%-------------------------------------------------------------------------------------------------
\subsection{Response to neutrons}
\label{sec:Neutrons}

\begin{figure}[htbp]
\centering
  \includegraphics[width=0.48\textwidth]{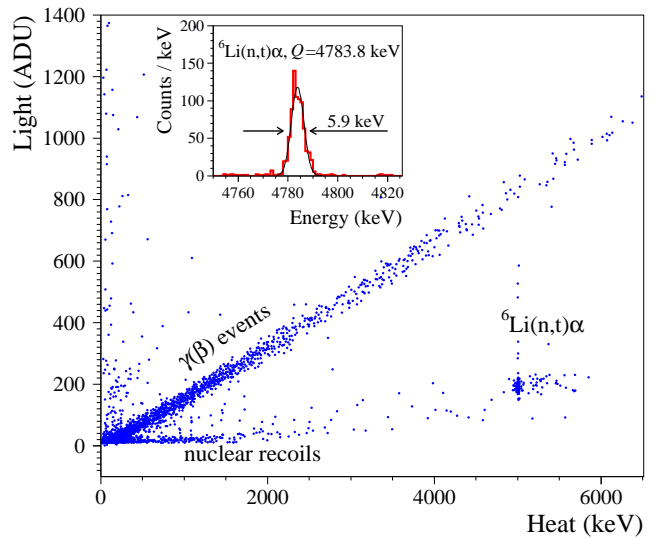}
\caption{The light-versus-heat data accumulated with the 186 g Li$_2$$^{100}$MoO$_4$ scintillating bolometer (enrLMO-t) in the EDELWEISS-III set-up (21--20 mK data) under neutron irradiation of 
an AmBe neutron source ($\approx 21$ n/s) over 33.5 h. Three populations ascribed to $\gamma$($\beta$)'s, $\alpha$+$^3$H events and nuclear recoils due to neutron scattering are well separated. 
(Inset) The $^6$Li thermal neutron capture peak, calibrated to the nominal energy of the reaction, together with a Gaussian fit. The energy resolution is FWHM = 5.9 keV}
\label{fig:enrLMO_L-vs-H}
\end{figure}

The ZMO-b, LMO-1, and enrLMO-t detectors were also exposed to neutrons from an AmBe source. The results for Li-containing bolometers are illustrated in Figs.~\ref{fig:enrLMO_L-vs-H} and \ref{fig:LMO_LY-vs-H} (left). The $\gamma$($\beta$) band exceeds the natural $^{208}$Tl end-point because of the prompt de-excitation $\gamma$'s following $^9$Be($\alpha$,n)$^{12}$C$^*$ reaction. The cluster of events in the $\alpha$ region is caused by the reaction $^6$Li(n,t)$\alpha$ ($Q$-value is 4784 keV \cite{ENDF}). The $^6$Li has a natural abundance of 7.5\% \cite{Meija:2016}, and the large cross section for thermal neutrons ($\sim$940 barns \cite{ENDF}) gives rise to the clear distribution at a heat energy of around 5 MeV. In the $\gamma$ energy scale, the distribution is shifted by about 7$\%$ with respect to the 4784~keV total kinetic energy released in the reaction. The energy resolution (FWHM) on the peak was measured as 7.7(3) and 5.9(2) keV for the LMO-1 and enrLMO-t detectors, respectively. This is an unprecedented result obtained with $^6$Li-containing detectors (e.g. compare with the results of Li-containing cryogenic detectors in Refs. \cite{Cardani:2013,Martinez:2012,Gironnet:2009} and references therein). A second structure at higher energy is attributed to the non-thermal neutrons, in particular to the resonant absorption of 240 keV neutrons. A linear fit to the less prominent lower band, ascribed to nuclear recoils induced by fast neutron scattering, gives a light yield of 0.07(2)~keV/MeV.

\nopagebreak

%-------------------------------------------------------------------------------------------------
\subsection{Particle identification by heat signals}
\label{sec:PSDheat}

As it was shown before, $\alpha$ particles exhibit a higher heat signal than $\gamma$($\beta$)s of the same energy. Even if a clear interpretation of this effect is lacking, this is probably related to the details of the phonon production mechanism in the particle interaction, which can lead to phonon populations with different features depending on the particle type. Therefore, one could expect some difference also in the shape of the heat signals between $\alpha$ and $\gamma$($\beta$) events and hence a pulse-shape discrimination capability of scintillating bolometers \footnote{And vice versa, the negligible, if any, difference in the thermal response to $\alpha$s and $\gamma$($\beta$)s, e.g. reported for TeO$_2$ \cite{Bellini:2010}, is probably responsible for the lack of a particle identification by the pulse-shape of non-scintillating bolometers, as it is the case of the TeO$_2$ bolometers.}. 

\nopagebreak
\begin{figure*}[htbp]
  \includegraphics[width=0.48\textwidth]{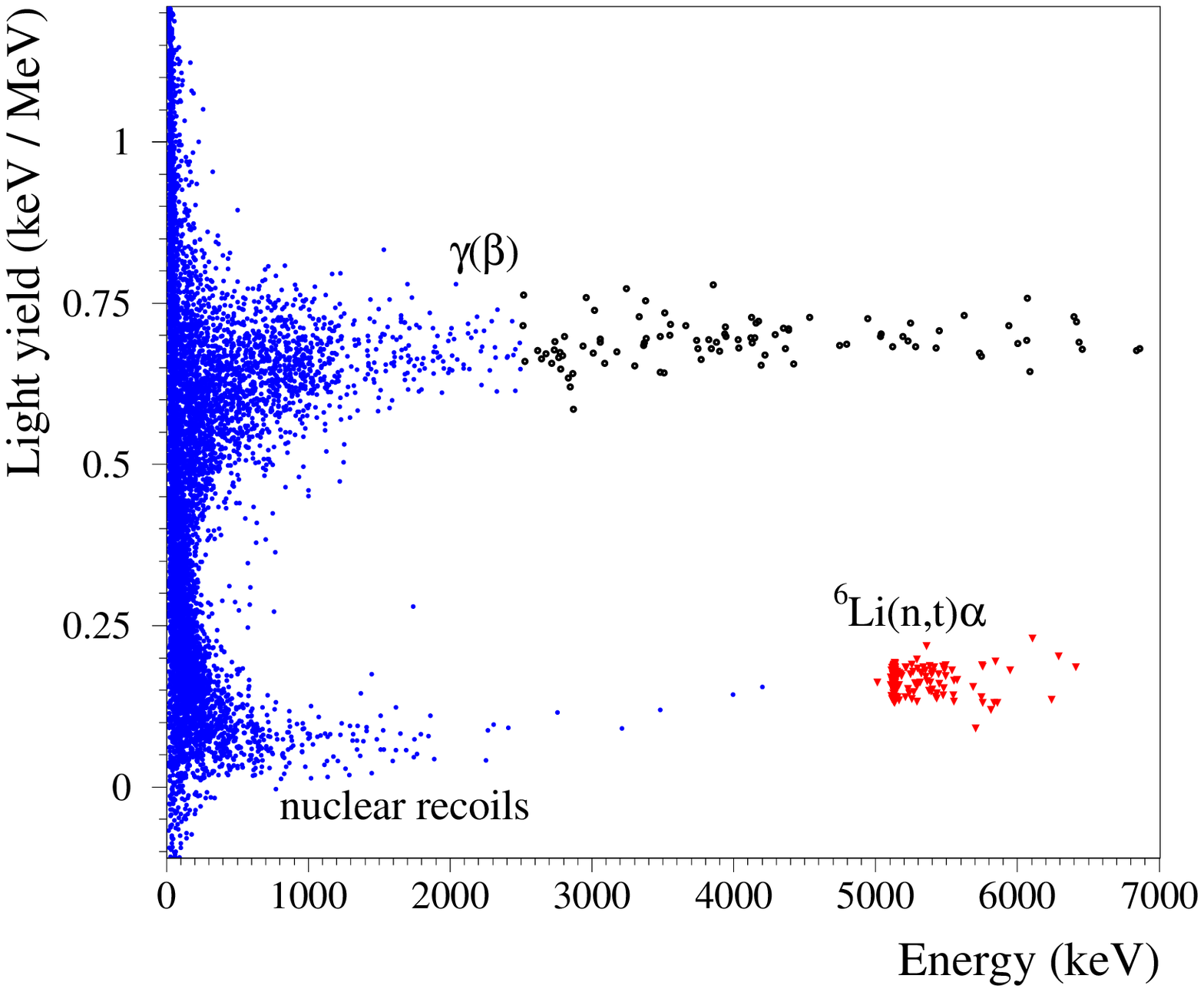}
  \includegraphics[width=0.48\textwidth]{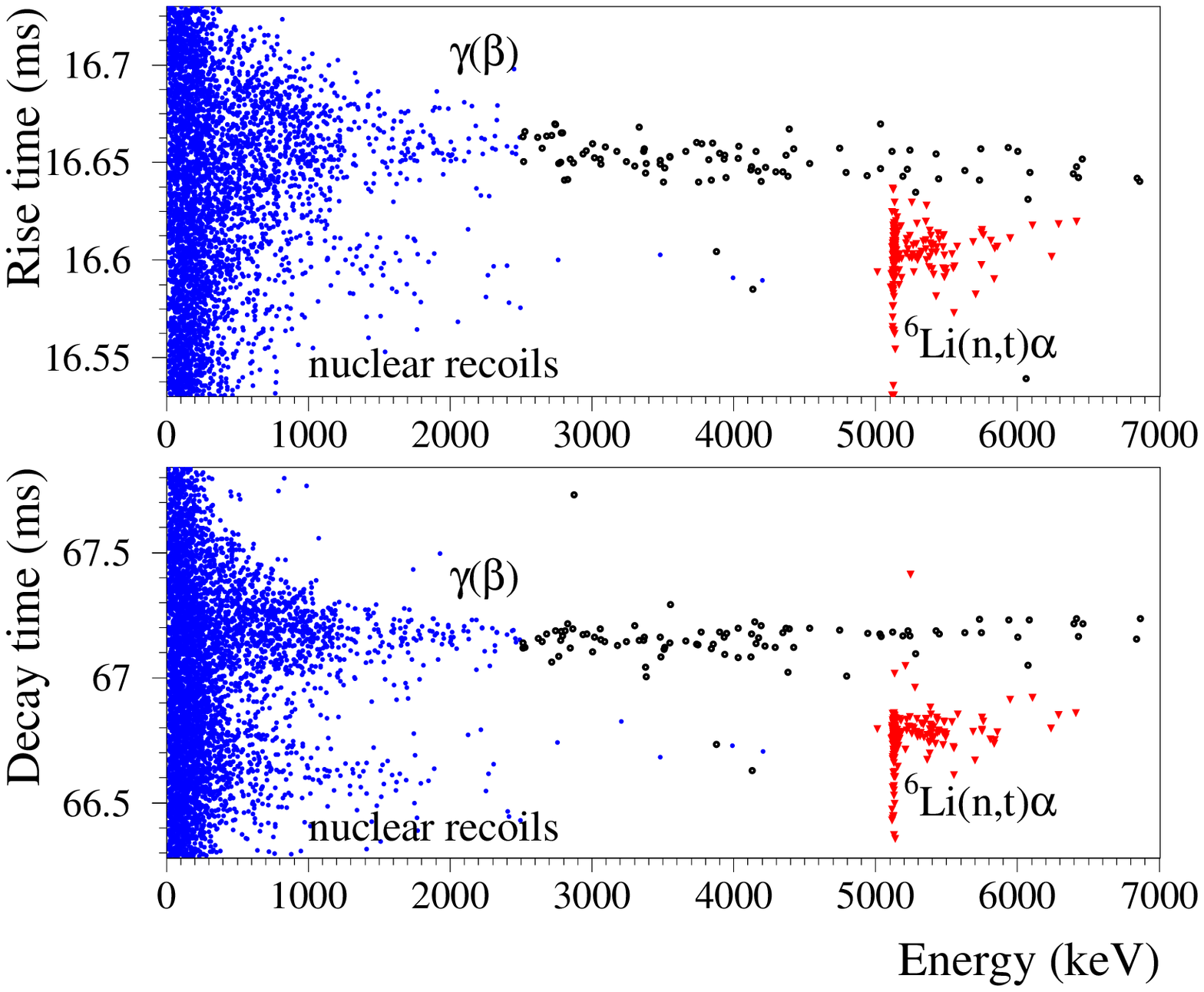}
\caption{Light-versus-heat scatter-plot obtained in a 20.5 h AmBe ($\sim $100 n/s) calibration measurement with a 151 g Li$_2$MoO$_4$ scintillating bolometer (LMO-1; left figure). Rise and decay times as functions of the energy (right figures). The populations of $\gamma(\beta)$ and $^6$Li(n,t)$\alpha$ events used for the evaluation of the discrimination power are marked by black cycles and red triangles, respectively. The calculated discrimination power is 19, 5.4, and 8.1 by means of the light yield, the rise time, and the decay time parameters, respectively}
\label{fig:LMO_LY-vs-H}
\end{figure*}

Previous measurements with ZnMoO$_4$ detectors de\-monstrated the possibility of pulse-shape discrimination by using only the heat channel \cite{Beeman:2012a,Gironi:2010,Beeman:2012b}. However, the discrimination ability strongly depends on the experimental conditions and sometimes can fail \cite{Cardani:2014}. No indication of this possibility has been claimed so far for Li$_2$MoO$_4$ bolometers.

A tiny difference between $\alpha$ and $\gamma$($\beta$) heat pulses of the enrZMO-t detector (about 3\% in the rising edge) allows us to perform an event-by-event particle identification using only the heat signals (e.g. $DP_{\alpha/\gamma(\beta)}$ = 3.8 was obtained for 2.5--3.5 MeV data). The data of the enrZMO-b bolometer, more affected by noise, show partial pulse-shape discrimination. It is worth noting that these results were obtained in spite of a low sampling rate (1 kSPS) and in one of the worst noise conditions among all the tested detectors.

Li$_2$MoO$_4$-based bolometers also demonstrate the possibility of the pulse-shape discrimination by 
a heat-signal shape analysis. Unfortunately, the data of most detectors were acquired with a low sampling rate (1~kSPS) and/or do not contain a large statistics of $\gamma(\beta)$ and $\alpha$ radiation in the same energy range, essential condition to investigate precisely this remarkable feature. However, significant results have been obtained by the analysis of the neutron calibration data (2~kSPS) of the LMO-1 detector. An example of the tiny difference in the time constants of $\gamma$($\beta$) and $\alpha$ heat pulses (less than $\sim$0.5 ms, i.e. a bin for the 2~kSPS sampling) is reported in Fig. \ref{fig:LMO_LY-vs-H} (right). By exploiting the rise and decay time parameters, we evaluated a $DP_{\alpha/\gamma(\beta)}$ between $\gamma(\beta)$s in the 2.5--7 MeV and $\alpha$-triton events in the 5--7 MeV range as 5.4 and 8.1, respectively. These results could probably be improved by using other pulse-shape parameters, as it was demonstrated with ZnMoO$_4$ detectors 
\cite{Beeman:2012a,Gironi:2010}. However, due to a few per mille difference of the thermal signals induced by $\gamma$($\beta$)s and $\alpha$s, the pulse-shape discrimination of scintillating bolometers is expected to be less efficient in comparison to the light-assisted particle identification which exploits an about 80\% difference in response (an exception for ZnMoO$_4$ has been reported in \cite{Beeman:2012a}). This is also the case for the LMO-1 detector, for which the double read-out allows to reach about twice better discrimination power. However, the requirement of 99.9\% rejection of $\alpha$-induced background (with a $\beta$ acceptance larger than 90\%) is achieved even for $DP_{\alpha/\gamma(\beta)}$ $\sim$ 3, therefore pulse-shape discrimination with the heat signals only could allow to simplify the detector structure and to avoid doubling the read-out channels in a CUPID-like $0\nu 2\beta$ experiment.

%-------------------------------------------------------------------------------------------------
\section{Backgrounds and radiopurity of $^{100}$Mo-containing scintillating bolometers}
\label{sec:Background}

%-------------------------------------------------------------------------------------------------
\subsection{Alpha background}
\label{sec:BKGalpha}

\begin{figure}[htbp]
\centering
\includegraphics[width=\columnwidth]{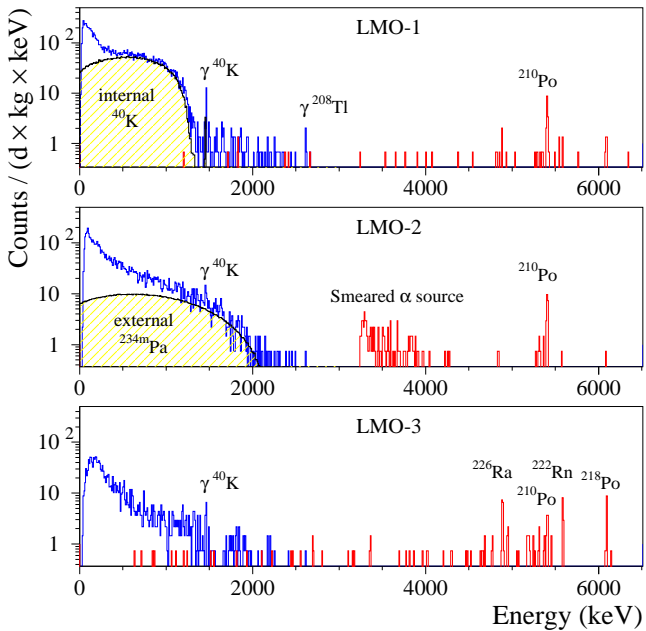}
\caption{The background energy spectra measured with the LMO-1 (over 237.5 h), LMO-2 (135 h), and LMO-3 (135 h) scintillating bolometers in the CUPID R\&D set-up. The energy bin is 10 keV. 
The $\alpha$ events in red are selected by the $LY$ parameter (the events of the $^{238}$U smeared $\alpha$ source for the LMO-2 detector are not shown below 3.25~MeV). An internal potassium contamination of the LMO-1 crystal generates the continuum up to $\approx1.3$~MeV and the $\gamma$ de-excitation peak at 1464 keV. The $^{208}$Tl line visible in the LMO-1 data can be ascribed to the thorium contamination of the set-up. The $\beta$ spectrum of $^{234m}$Pa in the data of the LMO-2 detector is due to the presence of the smeared $^{238}$U $\alpha$ source. The $\alpha$ peaks of $^{210}$Po (common for all the crystals) and $^{226}$Ra with daughters (in LMO-3) are caused by the contamination of the Li$_2$MoO$_4$ crystals}
\label{fig:LMOspectra}
\end{figure}

\begin{figure*}[htbp]
  \includegraphics[width=0.48\textwidth]{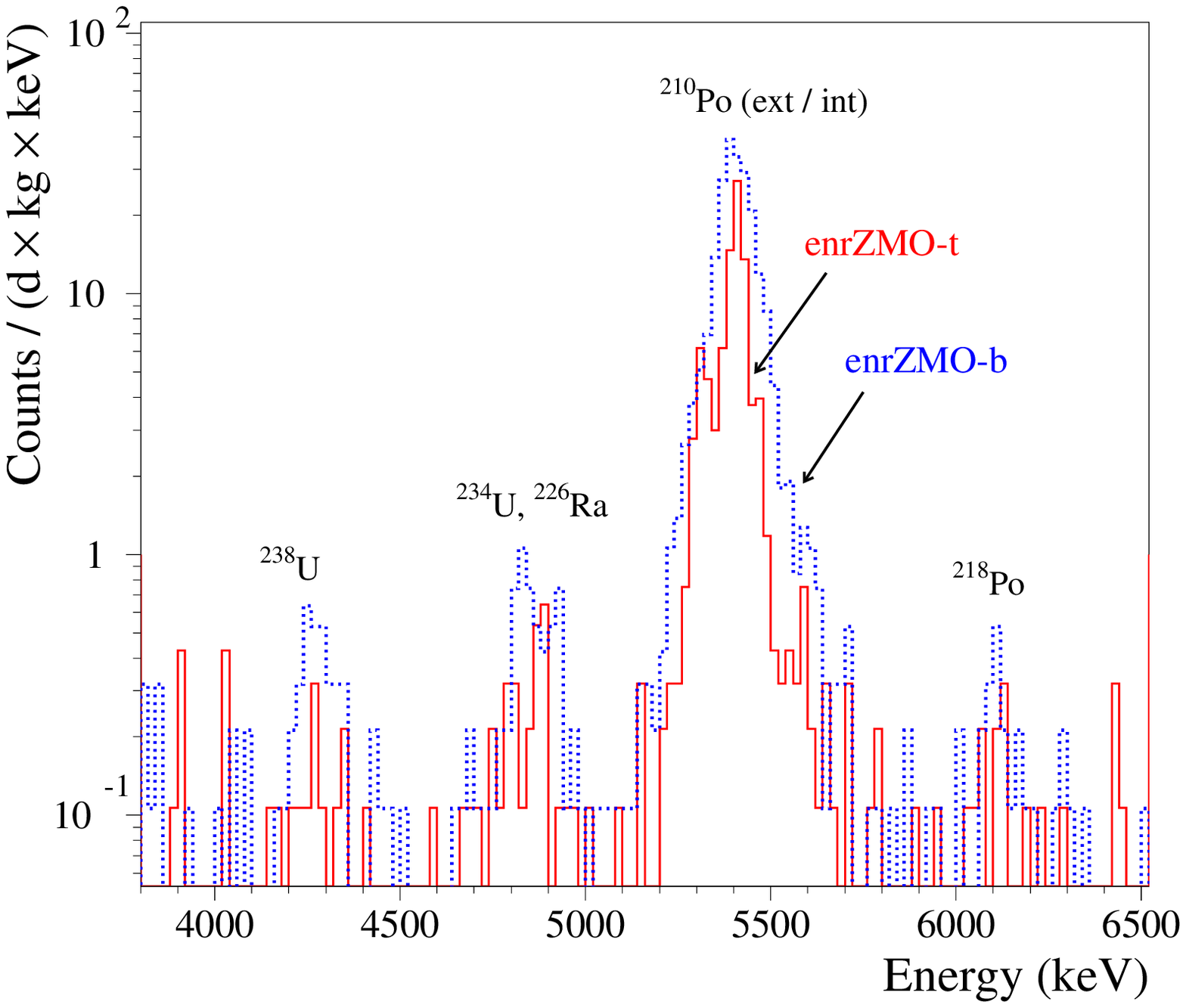}
  \includegraphics[width=0.48\textwidth]{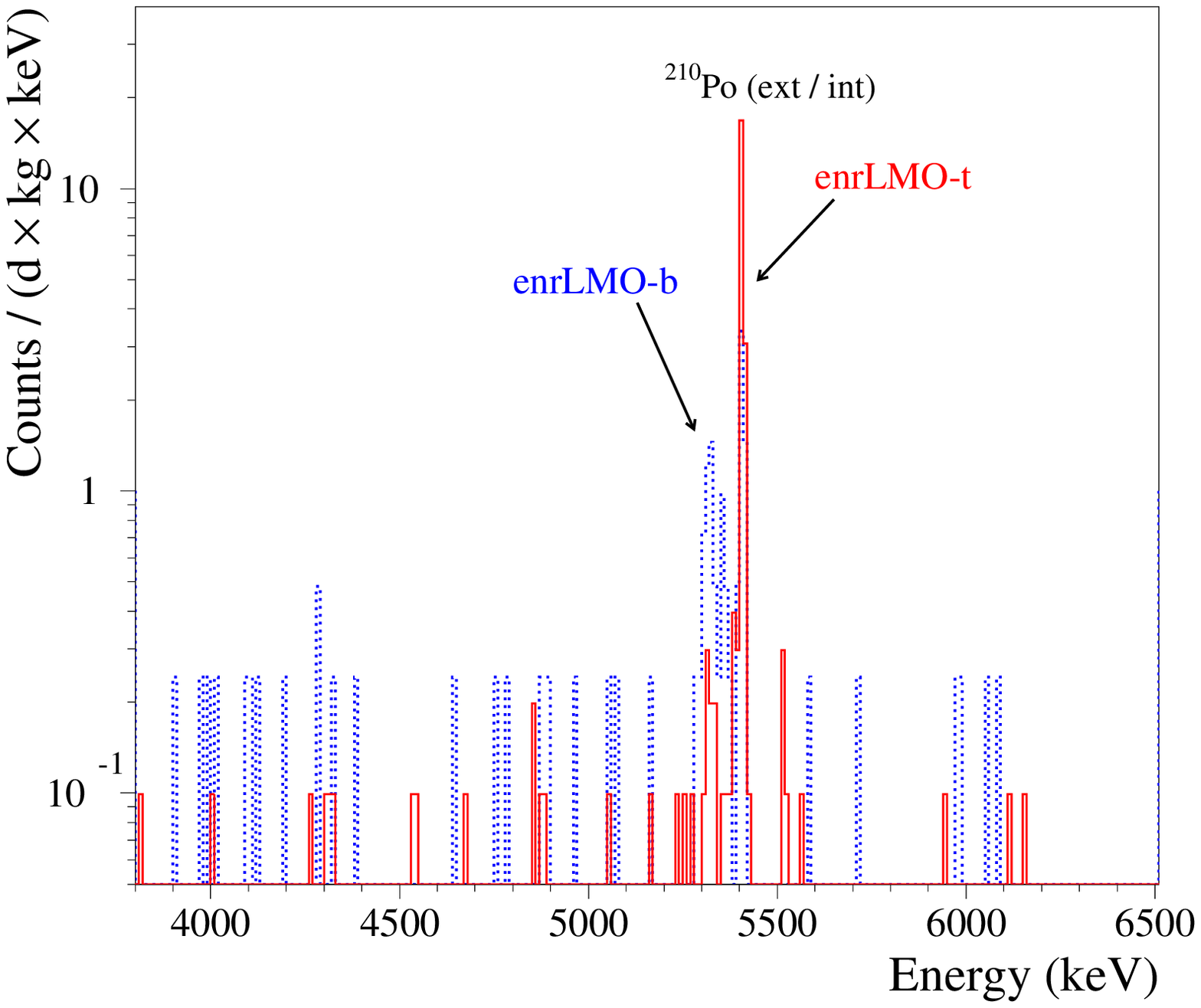}
\caption{The energy spectra of $\alpha$ events detected by the 0.4~kg Zn$^{100}$MoO$_4$ (left) and the 0.2~kg Li$_2$$^{100}$MoO$_4$ (right) scintillating bolometers. The energy bin is 20 keV and 10 keV, respectively. The data of enrZMO-t and enrZMO-b (both over 593 h), and enrLMO-b (487 h) detectors were collected in a low-background measurements in the CUPID R\&D cryostat at LNGS. The enrLMO-t (1303 h) bolometer has been operated in the EDELWEISS-III set-up at LSM}
\label{fig:enrDETspectra_a}
\end{figure*}

The $\alpha$ spectrum measured by the ZMO-b detector in Run308 can be found in \cite{Armengaud:2015,Poda:2015}, therefore the illustration of other spectra of the ZnMoO$_4$ bolometers is omitted. The background spectra of $\alpha$ events accumulated by the natural Li$_2$MoO$_4$ and all the enriched detectors are shown in Fig.~\ref{fig:LMOspectra} and Fig.~\ref{fig:enrDETspectra_a}, respectively. The anomaly (``dark hot $\alpha$'') in the response to $\alpha$s in the Zn$^{100}$MoO$_4$ bolometers was corrected by using the results of the fit to the $^{210}$Po events distribution in the $LY$-vs-heat data. The $^{232}$Th calibration data (168 h) of the enrLMO-b detector were combined with the background data to increase the statistics. 

All the crystals exhibit a contamination by $^{210}$Po, however we cannot distinguish precisely a surface $^{210}$Po pollution from a bulk one. Furthermore, most likely the observed $^{210}$Po is due to $^{210}$Pb contamination of the crystals, as this is the case for the ZnMoO$_4$ scintillator (ZMO-b) \cite{Armengaud:2015}. The LMO-3 crystal, produced from the Li$_2$CO$_3$ compound strongly polluted by $^{226}$Ra (see Table~\ref{tab:Li_purity}), is contaminated by $^{226}$Ra too. There is also a hint of a $^{226}$Ra contamination of the other natural ZnMoO$_4$ and Li$_2$MoO$_4$ crystals (ZMO-t, ZMO-b, LMO-1, and LMO-2), but the low statistics does not allow to estimate $^{226}$Ra activity in the crystals. In addition to $^{210}$Po and $^{226}$Ra, both Zn$^{100}$MoO$_4$ crystals demonstrate a weak contamination by $^{238}$U and $^{234}$U. 

\nopagebreak

The $\alpha$ spectra were analyzed to estimate the activity of $\alpha$ radionuclides from the U/Th chains and $^{190}$Pt. Determination of $^{190}$Pt activity in the detectors operated with the smeared $\alpha$ sources (enrZMO-t, enrZMO-b, and LMO-2) is difficult. We assumed that the energy resolution of the $\alpha$ peaks searched for is the same as the resolution of the $^{210}$Po peak present in the spectra of all detectors. The area of the $\alpha$ peaks was determined within $Q_{\alpha} \pm 3\sigma_{\alpha}$ energy interval, where $\sigma_{\alpha}$ is a standard deviation of the $^{210}$Po peak. If no peak observed, the Feldman-Cousins approach \cite{Feldman:1998} was applied to determine upper limits at 90\% C.L. A summary of the radioactive contamination of the natural and $^{100}$Mo-enriched ZnMoO$_4$ and Li$_2$MoO$_4$ crystals is given in Table \ref{tab:radiopurity}. 

The measured activity of $^{210}$Po in the crystals is $\sim$0.1--2~mBq/kg (see Table \ref{tab:radiopurity}). If the $^{210}$Po contamination of Li$_2$MoO$_4$ samples is originated by $^{210}$Pb, one can expect a growth of the $^{210}$Po activity up to the $\sim$1~mBq/kg. The limits on the activity of other radionuclides from the U/Th families have been set on the level of 0.001--0.05 mBq/kg (a few exceptions of contamination by $^{238,234}$U and/or $^{226}$Ra will be discussed below). A hint for a $^{190}$Pt content on the $\mu$Bq/kg level is evident only for the ZMO-b sample, while for other crystals it is below 4--11~$\mu$Bq/kg. It should be stressed that the sensitivity of the most measurements in the CUPID R\&D set-up at LNGS is limited by the low exposure, while the constrains on radioactive contamination of the ZMO-t crystal are affected by the vibrational noise induced poor energy resolution of the bolometer (e.g. FHWM $\sim 60$~keV at 5407~keV of $^{210}$Po). 

The efficient segregation of thorium and radium in the growing process is evident from the comparison of the radioactive contamination of the Li$_2$MoO$_4$ sample LMO-3 (Table \ref{tab:radiopurity}) and the Li$_2$CO$_3$ powder (Alfa Aesar in Table \ref{tab:Li_purity}) used for the crystal growth: the latter exhibits a clear pollution by $^{228}$Th (12 mBq/kg) and $^{226}$Ra (705 mBq/kg), while no indication of $^{228}$Th ($\leq$0.02 mBq/kg) and a significantly reduced activity of $^{226}$Ra (0.13 mBq/kg) were observed in the crystal. In addition, there is a clear sign of segregation of $^{238}$U and its daughters along the ZnMoO$_4$ crystal boule, because their concentration in the samples produced from the bottom part of the boule is around 2--4 times larger than that in the samples cut from the top part. Similar segregation has been reported, e.g., for CsI(Tl) \cite{Zhu:2006}, CaWO$_4$ \cite{Danevich:2011} and CdWO$_4$ \cite{Barabash:2011,Poda:2013,Barabash:2016}. The results of the present work and Refs.\cite{Velazquez:2017,Grigorieva:2017} show that the mechanism of segregation in the Li$_2$MoO$_4$ crystal growth process is less clear and further study would be useful to clarify this item. In general, it is expected \cite{Danevich:2011,Barabash:2016} that crystals produced by a double crystallization should be less contaminated. This is indeed observed for a Zn$^{100}$MoO$_4$ boule ($^{238}$U, $^{234}$U and $^{226}$Ra content at the level of few tens of $\mu$Bq/kg) in comparison to the recrystallized ZnMoO$_4$ and Li$_2$$^{100}$MoO$_4$ scintillators (only limits below ten $\mu$Bq/kg). In summary, it is evident that the radiopurity level of the $^{100}$Mo-enriched ZnMoO$_4$ and Li$_2$MoO$_4$ crystals satisfies the demands of a next-generation bolometric $0\nu 2\beta$ experiment \cite{Beeman:2012,Beeman:2012a,Artusa:2014a}.

%-------------------------------------------------------------------------------------------------
\subsection{Surface radioactive contamination}
\label{sec:BKGsurface}

As one can see in Fig. \ref{fig:light_vs_heat}, the counting rate of energy-degraded  $\alpha$s (that are expected due the surface contamination of the detector) in Run310 is lower than in previous runs. In particular, the $\alpha$ rate in the energy range 2.7--3.9~MeV, excluding the region of $^{190}$Pt, was reduced from 0.7(1) and 0.52(6)~counts/yr/kg/keV in Run308 (for ZMO-t and ZMO-b, respectively) to 0.20(6) and 0.09(5)~counts/yr/kg/keV in Run310 (for ZMO-b and enrLMO-t, respectively). These results are comparable to the 0.110(1) counts/yr/kg/keV rate measured in Cuoricino \cite{Andreotti:2011}), but a factor 5 worse than the purity achieved in CUORE-0 (0.016(1)~counts/yr/kg/keV \cite{Alfonso:2015}). However, it is worth noting that no special efforts were dedicated to surface cleaning in the ZnMoO$_4$ and Li$_2$MoO$_4$ detectors, while a significantly reduced amount of copper structure, a special surface treatment \cite{Alessandria:2013} and a dedicated mounting system \cite{Buccheri:2014} were adopted in CUORE-0.

%-------------------------------------------------------------------------------------------------
\subsection{Neutron background}
\label{sec:BKGneutron}

The data acquired with the Li$_2$$^{100}$MoO$_4$ detectors were used to estimate the thermal neutron flux inside the EDELWEISS-III and CUPID R\&D set-ups by exploiting the $\alpha$+t signature of neutron captures by $^6$Li. The data of the enrLMO-t detector do not contain any evidence of such events, while one event is found in the enrLMO-b data. The expected background in the region (the same used for the radiopurity analysis) is 0.054 (0.24) counts for enrLMO-t (enrLMO-b). According to Ref. \cite{Feldman:1998}, the number of events which can be excluded at 90\% C.L. is 2.39 (4.11) counts for the enrLMO-t (enrLMO-b). Assuming 100\% detection efficiency for such large-volume $^6$Li-containing detectors (e.g. see in Ref. \cite{Martinez:2012}) and taking into account the total live time (1303~h / 487~h) and the surface area (85.7~cm$^2$ / 90.4~cm$^2$) of the enrLMO-t / enrLMO-b crystals, we estimate the following upper limits on the thermal neutron flux inside the EDELWEISS-III and CUPID R\&D set-ups: 5.9$\times$10$^{-9}$ n/cm$^2$/s and 2.6$\times$10$^{-8}$~n/cm$^2$/s at 90\% C.L., respectively. The constraint for the EDELWEISS-III is comparable to the limit of $3\times10^{-9}$~n/cm$^2$/s reported in \cite{Rozov:2010} for the thermal neutron flux inside the lead/polyethylene shielding of the EDELWEISS-II set-up. The shield of both configurations of the set-up was the same except for 150 kg of polyethylene recently installed outside the thermal screens and inside the cryostat close to the detector volume. The limit for CUPID R\&D is by order of magnitude improved to the one, which can be extracted in the same way from the data of previous measurements with a 33 g Li$_2$MoO$_4$ scintillating bolometer in this set-up \cite{Cardani:2013}. It is worth noting that the deduced results are affected by an uncertainty which is difficult to estimate without the Monte Carlo simulations of the neutron propagation in the 
low temperature environment. It concerns a possible competition between the Li$_2$MoO$_4$ detectors and the neighbor materials in the capture of cold neutrons, further thermalized thermal neutrons as a result of interactions with a cold moderator of the set-up (e.g. 1 K polyethylene shield of the EDELWEIS-III). 

%-------------------------------------------------------------------------------------------------
\newpage
\clearpage

\begin{landscape}

\begin{table*}
\caption{Radioactive contamination of ZnMoO$_4$ and Li$_2$MoO$_4$ crystal scintillators. The errors of the activities are estimated at 68\% C.L., the upper limits are given at 90\% C.L. The $^{226}$Ra contamination of the LMO-3 sample is due to high activity of this radionuclide in the Li$_2$CO$_3$ powder used for the crystal growth (see text)}
%\footnotesize
\scriptsize
%\tiny
\begin{center}
\begin{tabular}{|c|c|ll|ll|lll|ll|}
 \hline
\multicolumn{2}{|c|}{Scintillator}  & \multicolumn{2}{c|}{\bf{ZnMoO$_4$}}    & \multicolumn{2}{c|}{\bf{Zn$^{100}$MoO$_4$}} & \multicolumn{3}{|c|}{\bf{Li$_2$MoO$_4$}} & \multicolumn{2}{c|}{\bf{Li$_2$$^{100}$MoO$_4$}} \\
\multicolumn{2}{|c|}{Mo sublimation} 			& \multicolumn{2}{c|}{Single} 	& \multicolumn{2}{c|}{Double}        & Single        & Single        & Single        & \multicolumn{2}{c|}{Double}  \\
\multicolumn{2}{|c|}{Mo recrystallization} & \multicolumn{2}{c|}{Double} 	& \multicolumn{2}{c|}{Double}        & Double        & Double        & Double        & \multicolumn{2}{c|}{Double}  \\
\multicolumn{2}{|c|}{Boule crystallization} 			& \multicolumn{2}{c|}{Double} 	& \multicolumn{2}{c|}{Single}        & Single        & Double        & Single        & \multicolumn{2}{c|}{Triple}  \\ 
\hline
\multicolumn{2}{|c|}{Crystal ID}           & ZMO-t  	& ZMO-b   & enrZMO-t	& enrZMO-b	& LMO-1	& LMO-2	& LMO-3      	& enrLMO-t	& enrLMO-b\\
\multicolumn{2}{|c|}{Position in boule}		& top    	& bottom  & top     	& bottom    & --    & --    & --      		& top   		& bottom\\ 
\multicolumn{2}{|c|}{Crystal mass} 				& 336 g  	& 334 g   & 379 g    	& 382 g     & 151 g	& 241 g & 242 g   		& 186 g 		& 204 g \\
\multicolumn{2}{|c|}{Radiputity test at}   & LSM			& LSM	 		& LNGS  		& LNGS  		& LNGS  & LNGS  & LNGS  			& LSM			 	& LNGS \\
\multicolumn{2}{|c|}{Time of measurements} & 1540 h		& 1300 h	& 593 h			& 593 h 		& 237 h & 135 h & 135 h 			& 1303 h		& 487 h \\
\hline
\bf{Activity}		& $^{232}$Th & $\leq7.3$ & $\leq1.4$ & $\leq8.0$ & $\leq9.0$ & $\leq18$ & $\leq21$ 	& $\leq18$ & $\leq2.7$		& $\leq11$ \\
\bf{($\mu$Bq/kg)} & $^{228}$Th & $\leq26$  & $\leq4.6$ & $\leq8.0$ & $\leq21$  & $\leq18$ & $\leq21$ 	& $\leq18$ & $\leq8.4$		& $\leq6.2$ \\
 \cline{2-11}
~           & $^{238}$U  & $\leq13$  & $\leq2.6$ & 10(4) 		& 39(7)     & $\leq18$ & $\leq37$ 	& $\leq48$ & $\leq4.9$		& $\leq11$ \\
~           & $^{234}$U  & $\leq20$  & $\leq3.0$ & 11(6)			& 43(10)    & $\leq18$ & $\leq21$ 	& $\leq46$ & $\leq6.7$		& $\leq11$ \\
~           & $^{230}$Th & $\leq28$  & $\leq1.4$ & $\leq17$  & $\leq24$  & $\leq18$ & $\leq21$ 	& $\leq18$ & $\leq2.7$		& $\leq11$ \\
~           & $^{226}$Ra & $\leq26$  & $\leq6.2$ & 14(3)     & 23(4)     & $\leq44$ & $\leq37$ 	& 130(19)  & $\leq6.7$		& $\leq11$   \\
~           & $^{210}$Po & 575(18) & 1320(30) & 809(32) & 2390(50) & 139(33) & 195(41) & 76(25) & 230(20) & 60(10) \\
\cline{2-11}
~           & $^{235}$U  & $\leq19$  & $\leq2.6$ & $\leq13$  & $\leq19$  & $\leq18$ & $\leq21$ 	& $\leq18$ & $\leq4.9$		& $\leq6.2$ \\
~           & $^{231}$Pa & -- 			  & $\leq1.4$ & $\leq8.0$ & $\leq36$  & $\leq18$ & $\leq21$ 	& $\leq18$ & $\leq2.7$		& $\leq6.2$ \\
~           & $^{227}$Th & -- 			  & $\leq2.6$ & $\leq8.5$ & $\leq12$  & $\leq18$ & $\leq21$ 	& $\leq18$ & $\leq4.9$		& $\leq6.2$ \\
\cline{2-11}
~           & $^{40}$K   & -- 				& --        & --          & --      & 62000(2000) & $\leq12000$ & $\leq3200$ & $\leq3500$	& $\leq3500$ \\
~           & $^{190}$Pt & $\leq4.4$ & 2.6(13)    & --          & --      & $\leq18$ & --        	& $\leq18$ & $\leq2.7$		& $\leq11$ \\
 \hline
\end{tabular}
\end{center}
\label{tab:radiopurity}
\end{table*}
\normalsize

\end{landscape}

%\clearpage
\newpage

%-------------------------------------------------------------------------------------------------
\subsection{Gamma(beta) background}

\subsubsection{$\gamma(\beta)$ background below 2615 keV}
\label{sec:BKGbelow2615}

\begin{figure}[htbp]
\centering
  \includegraphics[width=0.48\textwidth]{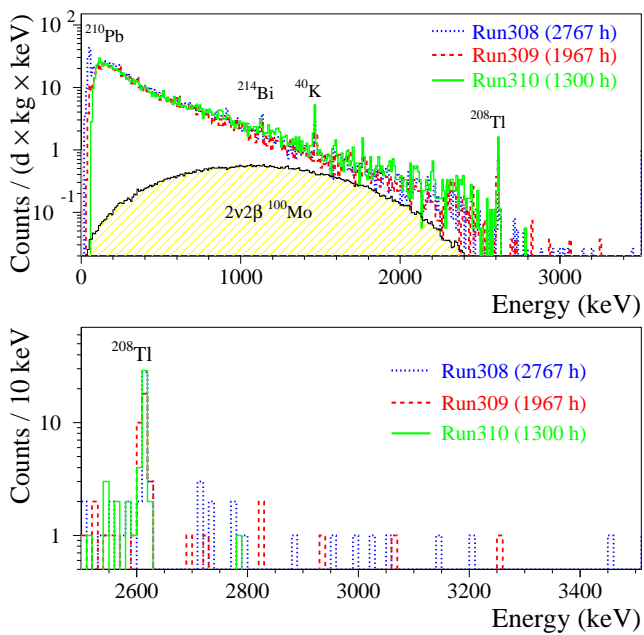}
\caption{The normalized energy spectra of $\gamma$($\beta$) events accumulated in low-background measurements with the 334 g ZnMoO$_4$ scintillating bolometer in the EDELWEISS-III set-up. 
A Monte-Carlo-simulated energy spectrum of the $2\nu2\beta$ decay of $^{100}$Mo with half-life $T_{1/2} = 6.90 \times 10^{18}$ yr (measured in the present work, see Sec.~\ref{sec:DBD}) is shown (upper panel). The energy bin is 10 keV. The same data in the 2.5--3.5 MeV energy interval (lower panel)}
\label{fig:ZMOspectra}
\end{figure}

The background spectra of $\gamma$($\beta$) events measured by the ZMO-b detector in the Runs 308--310 are shown in Fig.~\ref{fig:ZMOspectra}. The data acquired at different positions of the detector inside the cryostat are superimposed. Few of the $\gamma$ peaks present in the spectra are caused by the contamination of the set-up \cite{Scorza:2015} and the detector components by K, Th and U. The natural isotopic abundance of molybdenum contains the isotope of $^{100}$Mo at the level of 9.7\% \cite{Meija:2016}, therefore the $2\nu2\beta$ decay of this nucleus gives a dominant background above 1.5 MeV even for the non-enriched ZnMoO$_4$ detector (see Fig.~\ref{fig:ZMOspectra}).

\nopagebreak

The $\gamma$($\beta$) background accumulated by three Li$_2$MoO$_4$ detectors in the CUPID R\&D set-up is shown in Fig.~\ref{fig:LMOspectra}. The region below 1.5 MeV of the LMO-1 detector 
is dominated by $^{40}$K due to potassium contamination of the crystal. The main $^{40}$K decay mode (branching ratio BR = 89.3\% \cite{Wang:2017}) is a $\beta^-$ decay with $Q_{\beta}$ = 1311 keV 
\cite{Wang:2017a}. The 1460.8 keV de-excitation $\gamma$-quanta following the $^{40}$K electron capture in $^{40}$Ar$^{*}$ (BR = 10.7\%; K-shell electron binding energy is 3.2 keV) is also clearly visible with a total energy of 1464 keV. The $^{208}$Tl $\gamma$ peak in the LMO-1 data is due to the thorium contamination of the set-up. The $\gamma$($\beta$) spectra of the LMO-2 and LMO-3 detectors contain only the $^{40}$K peak caused by the potassium contamination of the set-up. The background of the LMO-2 bolometer is dominated by the $\beta$ spectrum of $^{234m}$Pa originated from the smeared $^{238}$U $\alpha$ source. The $^{40}$K activity in the LMO-1 and the limits for the LMO-2 and LMO-3 crystals are given in Table \ref{tab:radiopurity}. The comparison of the $^{40}$K content in the LMO-1 and LMO-2 crystals demonstrates a segregation of potassium in the crystal growth process by at least a factor of 5. 

\begin{figure*}[htbp]
  \includegraphics[width=0.48\textwidth]{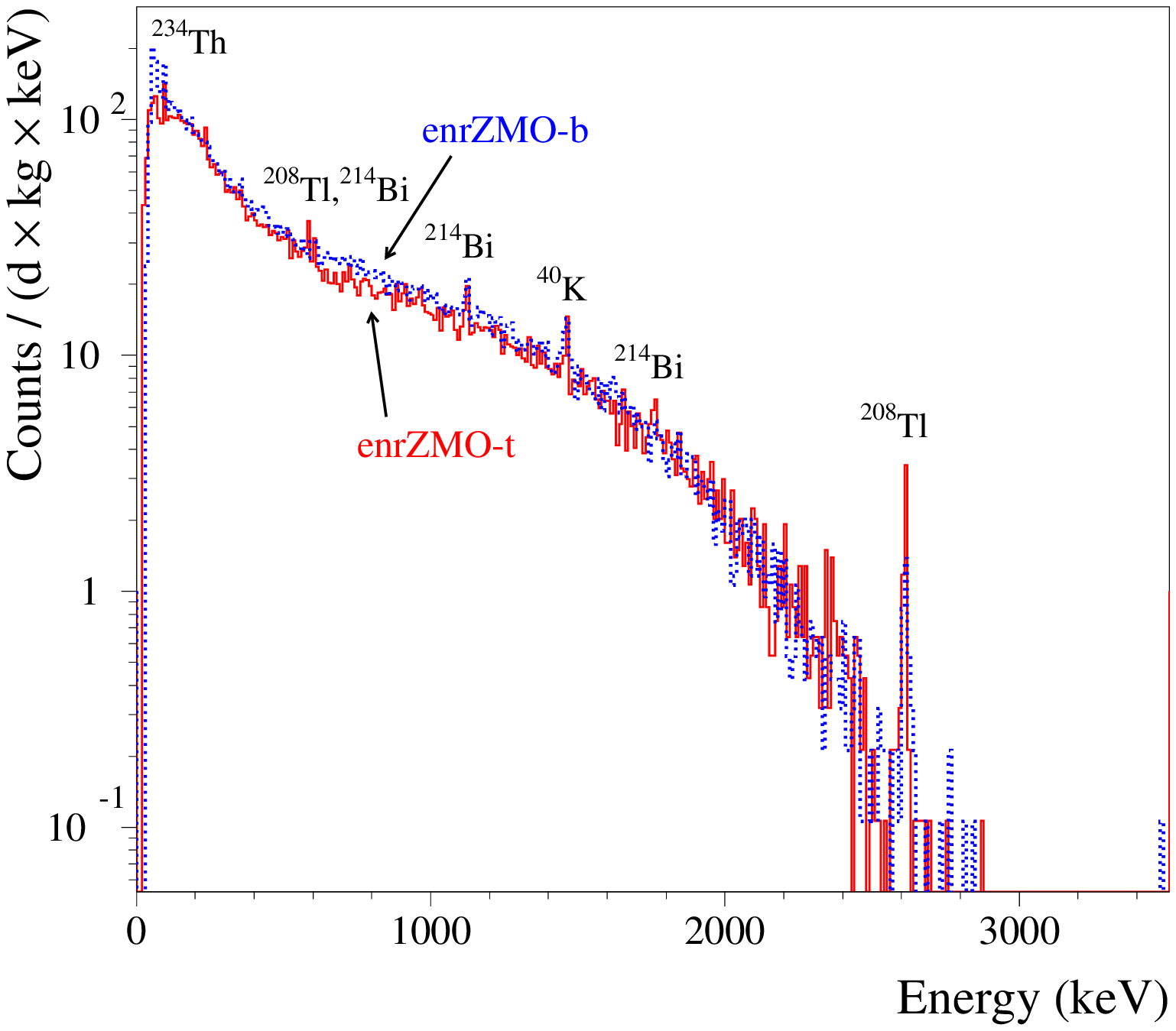}
  \includegraphics[width=0.48\textwidth]{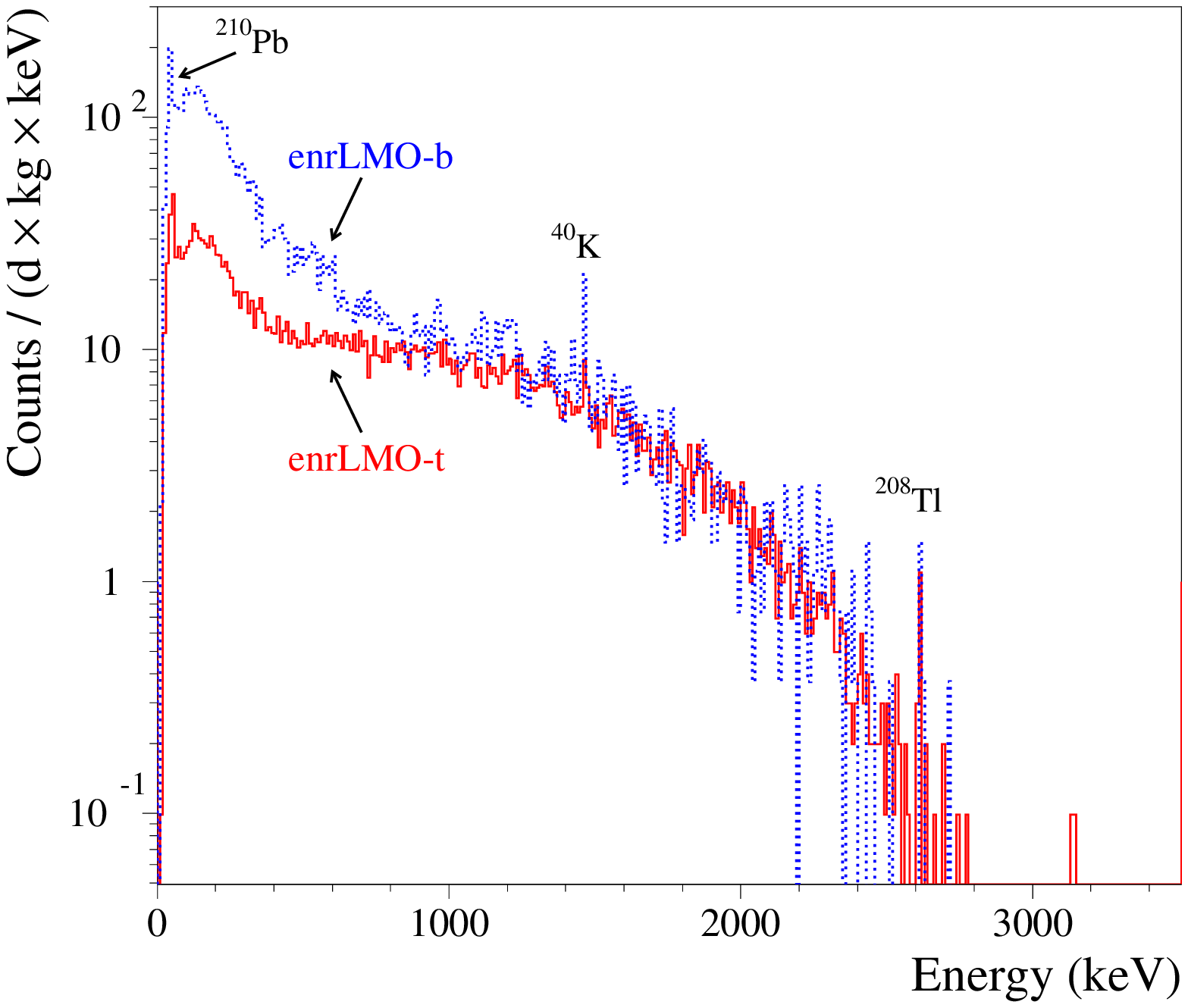}
\caption{The energy spectra of $\gamma$($\beta$) events measured by the $\sim 0.4$~kg Zn$^{100}$MoO$_4$ (left) and $\sim 0.2$~kg Li$_2$$^{100}$MoO$_4$ (right) scintillating bolometers. The energy bin is 10 keV. The data of both Zn$^{100}$MoO$_4$ detectors and one Li$_2$$^{100}$MoO$_4$ (enrLMO-b) detector were accumulated in the CUPID R\&D cryostat (593 h and 319 h of data taking, respectively), while the enrLMO-t bolometer was measured in the EDELWEISS-III set-up (over 1303 h). The origin of the most intensive $\gamma$ peaks is marked}
\label{fig:enrDETgamma}
\end{figure*}

Fig.~\ref{fig:enrDETgamma} shows the $\gamma$($\beta$) background of the $^{100}$Mo-enriched detectors dominated above 1 MeV by the $2\nu 2\beta$ decay of $^{100}$Mo with an activity of $\sim$10 mBq/kg. Some difference in the background counting rate for several $\gamma$ peaks measured by the Zn$^{100}$MoO$_4$ bolometers (Fig.~\ref{fig:enrDETgamma}, left) indicates a position-dependent background inside the CUPID R\&D set-up. In addition, one can see in Fig.~\ref{fig:enrDETgamma} (right) the excess of events below 0.8 MeV for the enrLMO-b data, which indicates a higher external background in the CUPID R\&D set-up in comparison to the EDELWEISS-III set-up. 

\nopagebreak

%-------------------------------------------------------------------------------------------------
\subsubsection{$\gamma(\beta)$ background above 2615 keV}
\label{sec:BKGabove2615}

The $\gamma(\beta)$ background spectra of the detectors (except natural Li$_2$MoO$_4$ samples) contain events above the 2615 keV $\gamma$ peak of $^{208}$Tl, see Fig.~\ref{fig:ZMOspectra} and \ref{fig:enrDETgamma}). An event-by-event analysis excludes that they are due to random coincidences. Also they cannot be explained by $\beta$ decay of $^{208}$Tl from the internal thorium contamination of the crystals, because the $^{228}$Th activity in the scintillators is low enough and no evidence of $^{212}$Bi $\alpha$ decays was found. The present surface purity of the detectors (see in Sec.~\ref{sec:BKGsurface}) does not play a role on the surface-induced $\gamma$($\beta$) background, whose contribution is expected to be about two orders of magnitude lower that that of the surface $\alpha$s \cite{Artusa:2014a}. These events can be originated by the muon-induced background, because no dedicated muon counter is available for the CUPID R\&D set-up, while the scintillating bolometers operated in the EDELWEISS-III did not have a synchronization with the available muon veto. However, at least for ZnMoO$_4$ detector ZMO-b the background above the 2615 keV $\gamma$ peak cannot be completely ascribed to muons because of a clear run-dependent difference in the counting rate of events in the 2.65--3.5~MeV energy range: 0.14(3), 0.08(3), and 0.02(2) counts/day in Runs 308, 309 and 310, respectively (see Fig.~\ref{fig:ZMOspectra}). The ZMO-b crystal was kept at sea level before Runs 308 and 310 over about 60 days. Therefore, cosmogenic activation, relevant for the tested crystals due to a rather short cooling period underground (typically, less than one month), cannot be the origin of the observed decrease of rate in time. A crucial difference in the ZnMoO$_4$ bolometer design of Run310 in comparison to early measurements is related to the absence of Mill-Max\textsuperscript{\textregistered} connectors with CuBe press-fit contacts that were previously placed on the external lateral surface of the detector holder. According to \cite{Scorza:2015}, a considerable part of the $\gamma$ background of the EDELWEISS-III set-up originates from radioactive contamination of the press-fit contacts (10(2) Bq/kg of $^{232}$Th; the total mass of the press-fit is 40 mg per connector). The Monte Carlo simulations of the connector-induced background of the ZMO-b bolometer show that 0.57\% of all decays of $^{208}$Tl populate the 2.65--3.5~MeV energy region, corresponding to a rate of 0.07~counts/day, comparable to the ones measured in Runs 308 and 309. Therefore, we can conclude that for the detectors tested in the EDELWEISS-III set-up the main source of $\gamma(\beta)$ events above 2615 keV is the detector's connectors. Moreover, as it is seen in Fig.~\ref{fig:enrDETgamma} (right), the $\gamma$($\beta$) background rate inside the CUPID R\&D set-up is even higher than that in the EDELWEISS-III, that can be explained by the radioactive contamination of the set-up. Therefore, special attention should be focused on selection of radiopure materials, in particular nearby the detectors, to realize a background-free $0\nu2\beta$ decay experiment.

%-------------------------------------------------------------------------------------------------
\subsubsection{Double-beta decay of $^{100}$Mo}
\label{sec:DBD}

\begin{figure}[htbp]
\centering
  \includegraphics[width=0.48\textwidth]{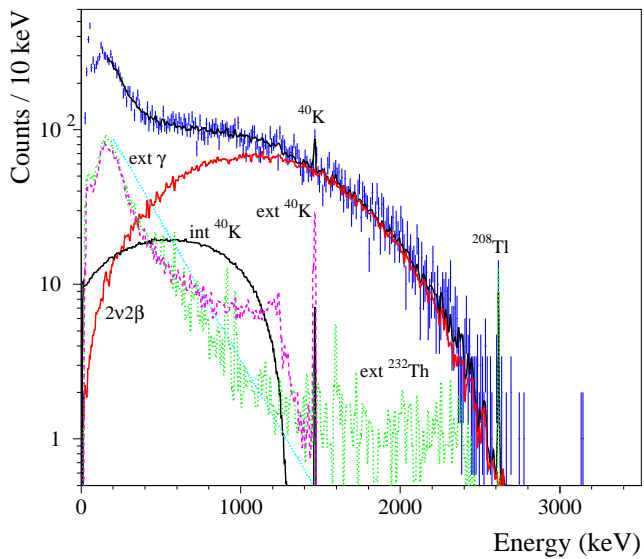}
\caption{The $\gamma$($\beta$) background spectrum accumulated over 1303 h with the 186 g Li$_2$$^{100}$MoO$_4$-based detector (enrLMO-t) in the EDELWEISS-III set-up together with the fit by a simplified background model built from the $2\nu 2\beta$ distribution of $^{100}$Mo ($T_{1/2}$ = 6.9$\times$10$^{18}$~yr), internal $^{40}$K (2.4~mBq/kg), and external $\gamma$ quanta represented by exponential background (ext $\gamma$), external $^{40}$K and $^{232}$Th. The $2\nu 2\beta$ signal-to-background ratio above 1.5~MeV is 8:1}
\label{fig:enrLMO_gamma}
\end{figure}

To extract the $^{100}$Mo $2\nu 2\beta$ decay half-life, the energy spectrum of the $\gamma$($\beta$) events accumulated by the enrLMO-t detector in the EDELWEISS-III set-up was fitted by a simplified background model (Fig. \ref{fig:enrLMO_gamma}). Taking into account a high crystal radiopurity, only two components of the internal background --- the $2\nu 2\beta$ decay of $^{100}$Mo and the bulk $^{40}$K decay --- are expected to give a significant contribution to the measured spectrum. The response function of the detector has been simulated with the help of the GEANT4-based code \cite{Simourg} and the DECAY0 event generator \cite{Decay0}. A model of the residual background (assuming it was caused by external $\gamma$ quanta from radioactive contamination of the materials surrounding the crystal) was built from an exponential function and the distributions of $^{40}$K and $^{232}$Th (built from the calibration data). The best fit ($\chi^2$/n.d.f. = 259.6/240 = 1.08) obtained in the energy interval 160--2700~keV gives (8853 $\pm$ 186)~decays of the $2\nu 2\beta$ of $^{100}$Mo and (1998 $\pm$ 605)~decays of the internal $^{40}$K. Taking into account the mass of the crystal (185.9 $\pm$ 0.1)~g, the live time (1303 $\pm$ 26)~h, and the pulse-shape discrimination efficiency (97.1 $\pm$ 0.4)\%, the bulk activity of $^{40}$K in the enrLMO-t crystal is estimated to be (2.4 $\pm$ 0.7)~mBq/kg (or $\leq$3.5~mBq/kg). The enriched crystal ((96.9 $\pm$ 0.2)\% of $^{100}$Mo) contains 6.103$\times$10$^{23}$ nuclei of $^{100}$Mo and, therefore, the $2\nu 2\beta$ decay half-life of $^{100}$Mo is $T_{1/2}$ = (6.90 $\pm$ 0.15)$\times$10$^{18}$ yr (statistical uncertainty only). Systematic uncertainties are related to the Monte Carlo simulations (5\% --- corresponds to the uncertainty of the GEANT4 modeling of electromagnetic interactions \cite{Amako:2010}) and to the fit by the background model (0.4\%). To estimate the latter value, we made a fit with the described model in different energy intervals: the left side was varied in the 160--300~keV range with a 10~keV step, while the right side was varied within the interval 2650--2750~keV with a 50~keV step. The fits gave a number of $^{100}$Mo $2\nu 2\beta$ decays in the range of 8821--8888. The combination of all the contributions results 5.4\% as a total systematic uncertainty. So, the half-life of the $2\nu 2\beta$ decay of $^{100}$Mo is measured to 

\begin{center}
$T_{1/2}$ = $\left[6.90 \pm 0.15(\mathrm{stat.}) \pm 0.37(\mathrm{syst.})\right] \times 10^{18}$ yr.
\end{center}

\nopagebreak

The obtained value is in a good agreement with the most accurate results achieved by NEMO-3 experiment, $\left[7.11 \pm 0.02(\mathrm{stat.}) \pm 0.54(\mathrm{syst.})\right] \times 10^{18}$~yr \cite{Arnold:2005} and [6.93 $\pm$ 0.04(stat.)] $\times$ 10$^{18}$~yr \cite{Arnold:2015}\footnote{The value of Ref. \cite{Arnold:2015} is a preliminary result of the NEMO-3 Phase I+II obtained from the fitting to the data above 2~MeV; the complete analysis is ongoing.}, and the bolometric measurements with ZnMoO$_4$ crystals 
[7.15 $\pm$ 0.37(stat.) $\pm$ 0.66(syst.)] $\times$ 10$^{18}$~yr \cite{Cardani:2014}, as well as with the average value $\left[7.1 \pm 0.4\right] \times 10^{18}$~yr \cite{Barabash:2015}. 

Because of the large mass and the relatively long measurement ($\sim$600~h), the largest exposure was accumulated with the two $\sim$0.4~kg Zn$^{100}$MoO$_4$ detectors enrZMO-t and enrZMO-b operated 
in the CUPID R\&D set-up at the Gran Sasso laboratory, containing $\sim 2 \times 10^{24}$ $^{100}$Mo nuclei. The usage of the smeared $^{238}$U source, which also emits electrons with an end-point $\sim 2$~MeV (see above the case of the LMO-2 detector), prevents us from getting a more precise $2\nu 2\beta$ half-life value of $^{100}$Mo than that obtained from the analysis of the enrLMO-t background. However, these data were used to search for neutrinoless double-beta decay of $^{100}$Mo and no counts were observed in the region of interest around 3034~keV. Considering an efficiency of $\sim$ 75\% in a 10~keV energy window, we set a limit on $0\nu2\beta$ decay of $^{100}$Mo of $2.6 \times 10^{22}$~yr at 90\% C.L. Of course, this result is by far inferior to that achieved by NEMO-3 with 6.914~kg of $^{100}$Mo over the live time of 4.96~yr ($T_{1/2} \geq 1.1 \times 10^{24}$ yr at 90\% C.L. \cite{Arnold:2015}), but --- given the low sensitive mass and the short duration of the test --- it shows the high potential of scintillating bolometers approach.

%################################################################################################
\section{Down selection of $^{100}$Mo-based scintillating bolometers technology}
\label{sec:Selection}

According to the results of the present work and some early related investigations, the state-of-the-art of the ZnMoO$_4$ and Li$_2$MoO$_4$ scintillating bolometer technology is summarized below:

\begin{itemize}
	
	\item In spite of about 10\% higher concentration of molybdenum (55\% vs. 44\% weight), the unit volume of 	Li$_2$MoO$_4$ contains $\sim$5\% less Mo because of the lower density (3.04 g/cm$^3$ vs. 4.18 g/cm$^3$; see the properties of the materials in \cite{Bekker:2016} and \cite{Berge:2014} respectively).
	
	\item Naturally occurring Zn and Li do not contain radioactive isotopes. The only radioactive isotope in Mo natural composition is $^{100}$Mo itself and its comparatively ``fast'' $2\nu 2\beta$ decay rate ($\sim$10 mBq activity in the enriched crystal) requires fast detector's response and pulse shape discrimination to avoid populating the $0\nu 2\beta$ decay ROI of $^{100}$Mo by $2\nu 2\beta$ pile-uped events.	
	
	\item A rather low  melting point (705 $^{\circ}$C vs. 1003 $^{\circ}$C) and the absence of phase transitions are compatible with comparatively easier Li$_2$MoO$_4$ crystallization with respect to ZnMoO$_4$, and lower losses of the enriched material during the crystal growth process (0.1\% vs. 0.6\%). 
	
	\item Highly purified $^{100}$Mo-enriched molybdenum oxide \cite{Berge:2014} is usable for both materials. No special purification is needed for use with commercially available high purity zinc oxide and lithium carbonate. 	However, there is an issue with high $^{40}$K contamination of Li-containing powder due to chemical affinity 	of lithium and potassium. Therefore, pre-screening measurements and purification are required to reduce potassium contamination in the crystal scintillators.
	
	\item Double crystallization is an efficient approach to produce high optical quality radiopure ZnMoO$_4$ and Li$_2$MoO$_4$ crystal scintillators.
	
	\item The established technology of Li$_2$MoO$_4$ crystal growth allows the use of most of the material for the production of scintillation elements. In a case of ZnMoO$_4$, the crystalline material quality along the boules is not stable enough to reach the same high level of the ready-to-use scintillation elements production.
	
	\item The hygroscopicity of Li$_2$MoO$_4$ is weak enough not to require a strict handling for the production of scintillation elements, mounting and operation of the detectors. The necessity of further improvement of the crystal surface purity is not presently evident but it would require the development of special mechanical/chemical treatment. ZnMoO$_4$ is not hygroscopic and therefore an acid etching could be applied to improve the surface purity if it is needed.
	
	\item The time response of ZnMoO$_4$ and Li$_2$MoO$_4$ bolometers equipped with an NTD Ge thermistor (order of one to few tens ms) is comparable with an efficient suppression of the background caused by pile-ups of the $2\nu 2\beta$ decay of $^{100}$Mo. A further improvement (below 10$^{-4}$~counts/yr/kg/keV) is expected with faster temperature sensors, e.g. Metallic Magnetic Calorimeters \cite{Luqman:2017}, or with light detectors with enhanced signal-to-noise ratio (e.g. exploiting the signal amplification by Neganov-Luke effect \cite{Chernyak:2017}).
	
	\item The energy resolution of Li$_2$MoO$_4$ bolometers satisfies the CUPID requirement. The resolution of ZnMoO$_4$ does not meet this requirement by a factor of 2. Moreover, an addition degradation by a factor of 2 is observed for the $^{100}$Mo-enriched ZnMoO$_4$ crystals produced from the bottom part of the boule with presently available quality;
	
	\item Typical light yield of Li$_2$MoO$_4$ is about 30\% lower than that of ZnMoO$_4$. However, light-assisted alpha rejection at satisfactory high level (8$\sigma$ and more) is achieved by detectors based on both materials.
	
	\item The imperfections of ZnMoO$_4$ crystals affect the bolometric response to bulk $\alpha$ events, a fraction of which is characterized by more quenched light and enhanced thermal signals. The observed anomaly does not spoil the $\alpha$ rejection capability, but affects the quality of the $\alpha$ spectroscopy. This is not an issue of Li$_2$MoO$_4$ bolometers thanks to a significantly higher crystal's quality.
	
	\item The ability to perform heat-pulse-shape discrimination is a feature of both Li$_2$MoO$_4$ and ZnMoO$_4$ detectors which allows a substantial simplification of the detector structure. The reproducibility of alpha particles rejection at the level of about 3$\sigma$ has to be demonstrated but it is not not mandatory for the scintillating bolometer technique.
		
	\item High thermal-neutron cross section of $^{6}$Li ($\sim$1 kb) leads to $^{6}$Li(n,t)$\alpha$ reaction, which can be exploited to suppress neutron-induced background. The lack of a similar feature in ZnMoO$_4$ would not suppress such background due to (n,$\gamma$) reactions on Zn, Mo, and O isotopes which produce $\gamma$ quanta with energies up to 7 MeV \cite{CapGam}.
	
	\item A possible cosmogenic activation of Li$_2$MoO$_4$ is expected to be much less significant than that of ZnMoO$_4$ because no cosmogenically activated isotopes, with the decay energy high enough to contribute to the ROI, can be produced from lithium natural isotopes (in contrast to zinc isotopes). Therefore, a cosmogenic-originated background of Li$_2$MoO$_4$ would be only associated to molybdenum.
	
	\item Very low contamination of both materials by U/Th completely satisfies the radiopurity demands even in the case of a single crystallization. The second crystallization further improves the crystals radiopurity thanks to the observed segregation of radioactive impurities. 		
\end{itemize}

In conclusion, the advanced crystal production process and better detector performance are crucial advantages of Li$_2$MoO$_4$ with respect to ZnMoO$_4$. For these reasons, Li$_2$MoO$_4$ was selected for the realization of a 10-kg-scale $0\nu 2\beta$ experiment (CUPID-0/Mo) aiming at demonstrating the viability of the LUMINEU scintillating bolometer technology for CUPID. Mass production of twenty enriched crystals with a size of $\oslash$44$\times$45~mm has been recently completed for the first phase of this experiment, to be performed in the EDELWEISS set-up at LSM (France). The start of CUPID-0/Mo phase-I data taking is planned in early 2018 and the full-scale operation is expected by the end of the year. A second phase will follow aiming at a full use of the available 10~kg of $^{100}$Mo.

%################################################################################################
\section{Conclusions}

A technology suitable for mass production of massive ($\sim$1~kg), high-optical-quality zinc and lithium molybdate crystal scintillators from highly purified molybdenum enriched in $^{100}$Mo has been established. The required performance and radiopurity of scintillating bolometers based on large-volume (50--90 cm$^3$) ZnMoO$_4$ and Li$_2$MoO$_4$ crystals (including $^{100}$Mo-enriched) 
have been demonstrated in low-background measurements at the Modane and Gran Sasso underground laboratories. 

The detectors show an excellent energy resolution (in particular, 4--6 keV FWHM of Li$_2$MoO$_4$ detectors at the 2615 keV $\gamma$ quanta of $^{208}$Tl), which is among the best resolution ever achieved with massive bolometers. The exploited heat-light dual read-out provides an efficient particle discrimination between $\gamma$($\beta$) and $\alpha$ events, which is compatible with more than 99.9\% $\alpha$ rejection while preserving approximately 100\% selection efficiency of a $0\nu2\beta$ signal. Furthermore, we demonstrated the possibility of pulse-shape discrimination by using the heat channel only, which is an important step towards detector simplification for the CUPID experiment. The Li$_2$MoO$_4$ scintillating bolometers are also found to be excellent neutron low-counting detectors. Their operation in the Modane and Gran Sasso cryogenic set-ups has allowed us to set very stringent limits, on the level of $\sim 10^{-8}$ neutrons/cm$^2$/s, on the thermal neutron flux in the EDELWEISS-III and 
CUPID R\&D facilities.

The radioactive contamination of the developed $^{100}$Mo-enriched crystal scintillators is very low. The activity of $^{232}$Th ($^{228}$Th) and $^{226}$Ra is below 10 $\mu$Bq/kg (down to a few 10 $\mu$Bq/kg of $^{226}$Ra in case of single boule crystallization). The total bulk $\alpha$ activity of U/Th is below a few mBq/kg. The activity of $^{40}$K in the Li$_2$$^{100}$MoO$_4$ samples is less than 4 mBq/kg. 
The $\gamma$($\beta$) background of the enriched detectors is dominated by the $^{100}$Mo $2\nu2\beta$ decay with a $\sim$10 mBq/kg activity.

By utilizing the data accumulated over about 50~days with a 0.2~kg Li$_2^{100}$MoO$_4$ detector, the half-life of $^{100}$Mo relative to the $2\nu2\beta$ decay to the ground state of $^{100}$Ru is measured with up-to-date highest accuracy: $T_{1/2}$ = $\left[6.90 \pm 0.15(\mathrm{stat.}) \pm 0.37(\mathrm{syst.})\right] \times 10^{18}$ yr. The sensitivity to $0\nu2\beta$ decay half-life on the order of 10$^{22}$~yr has been reached with 0.8~kg $^{100}$Mo-enriched detectors operated over less than one month. It is far to be competitive to the most stringent limits in the field (e.g. $\sim$10$^{24}$~yr limit deduced by NEMO-3 for $^{100}$Mo), but this sensitivity was achieved by using 3--4 orders of magnitude less accumulated statistics in comparison with the leading $0\nu2\beta$ experiments. These results definitely demonstrate a potential of scintillating bolometers to perform high sensitivity $2\beta$ searches.

Taking into account the reproducible technology to grow large-mass, high-optical quality Li$_2$MoO$_4$ crystals and their high bolometric performance together with low radioactive contamination, Li$_2$MoO$_4$-based scintillating bolometers have been chosen for the realization of a cryogenic $2\beta$ experiment with $\approx$10 kg of enriched $^{100}$Mo (CUPID-0/Mo) to prove the viability of this approach for CUPID. The first batch of twenty 0.2-kg Li$_2$$^{100}$MoO$_4$ crystal scintillators has been produced to carry out a first phase of the experiment in the EDELWEISS-III set-up at Modane (France).

%-------------------------------------------------------------------------------------------------
\begin{acknowledgements}
This work is a part of the program of LUMINEU, a project receiving funds from the Agence Nationale de la Recherche (ANR, France). The development of Li$_2$MoO$_4$ scintillating bolometers was 
carried out within ISOTTA, a project which received funds from the ASPERA 2$^{nd}$ Common Call dedicated to R\&D activities. This work has been funded in part by the P2IO LabEx (ANR-10-LABX-0038) 
in the framework ``Investissements d'Avenir" (ANR-11-IDEX-0003-01) managed by the Agence Nationale de la Recherche (France), by the European Research Council (FP7/2007-2013) under contract LUCIFER 
no. 247115 and by the Italian Ministry of Research under the PRIN 2010-2011 contract no. 2010ZXAZK9. The group from Institute for Nuclear Research (Kyiv, Ukraine) was supported 
in part by the IDEATE International Associated Laboratory (LIA), and by the project ``Investigation of neutrino and weak interaction in double beta decay of $^{100}$Mo'' in the framework of the Programme ``Dnipro'' based on Ukraine-France Agreement on Cultural, Scientific and Technological Cooperation. F.A.D. acknowledges the support of the Universit\'{e} Paris-Sud. A.S.Z. is supported by the ``IDI 2015'' project funded by the IDEX Paris- Saclay, ANR-11-IDEX-0003-02. The help of the technical staff of the Laboratoire Souterrain de Modane and the participant laboratories is gratefully acknowledged. 
We thank the mechanical workshop of CEA/SPEC for its skillful contribution to the conception and fabrication of detector holders. The authors are grateful to the reviewers, as well as to Joesph~Johnston and Alexander~Leder (both from the Massachusetts Institute of Technology), for a careful reading of the manuscript, helpful corrections, and valuable comments.

\end{acknowledgements}

%--------------------------------------------------------------------------------
%\begin{thebibliography}{9}

\end{document}